\documentclass[final,3p,times,twocolumn]{elsarticle}
\usepackage[utf8]{inputenc}

\usepackage{lipsum} 
\usepackage{xcolor} 
\usepackage{amsmath} 
\usepackage{hyperref} 
\usepackage{subcaption} 
\usepackage{url}
\usepackage{bm}
\usepackage{placeins} 
\usepackage{stfloats}
\usepackage{mwe}
\usepackage{array,multirow} 
\usepackage{mathrsfs}  

\usepackage[mathlines]{lineno} 

\title{Topology Optimization of Two Fluid Heat Exchangers}
\author[add1]{Lukas Christian Høghøj\corref{cor1}}
\ead{luch@mek.dtu.dk}
\author[add1]{Daniel Ruberg Nørhave}
\author[add2]{Joe Alexandersen}
\author[add1]{Ole Sigmund}
\author[add1]{Casper Schousboe Andreasen}
\cortext[cor1]{Please address correspondence to luch@mek.dtu.dk}
\address[add1]{Department of Mechanical Engineering, Section for Solid Mechanics, Technical University of Denmark, Kgs. Lyngby, Denmark}
\address[add2]{Department of Technology and Innovation, University of Southern Denmark, Odense, Denmark}
\date{June 2020}
\journal{arXiv.org}

\newcommand{\diff}[2]{\frac{\partial#1}{\partial#2}}
\newcommand{\ddiff}[2]{\frac{\mathrm{d}#1}{\mathrm{d}#2}}

\renewcommand{\Re}{\mathit{Re}}
\newcommand{\Pe}{\mathit{Pe}}

\newcommand{\Pes}{\mathit{Pe}_s}
\newcommand{\Ck}{C_k}

\newcommand{\x}{\mathbf{x}}

\newcommand{\xiI}{\xi_1}
\newcommand{\xihtI}{\hat{\tilde{\xi}}_1}
\newcommand{\xiII}{\xi_2}
\newcommand{\xihtII}{\hat{\tilde{\xi}}_2}
\newcommand{\xiht}{\hat{\tilde{\xi}}}
\newcommand{\xip}{\hat{\tilde{\xi}}}
\newcommand{\xih}{\hat{\xi}}
\newcommand{\xit}{\tilde{\xi}}

\newcommand{\amax}{\overline{\alpha}}
\newcommand{\abmax}{\overline{\alpha}^\gamma}
\newcommand{\amI}{\overline{\alpha}^1}
\newcommand{\amII}{\overline{\alpha}^2}

\newcommand{\aI}{\alpha^1}
\newcommand{\aII}{\alpha^2}

\newcommand{\phih}{\Phi_\mathit{coolant}}

\newcommand{\dpmax}{\Delta P_\mathit{max}}
\newcommand{\dpbase}{\Delta P_\mathit{base}}
\newcommand{\dpf}[1]{\ensuremath{\dpmax=#1\dpbase}}
\newcommand{\obj}[1]{\ensuremath{\phih=#1}}

\newcommand{\TPS}{\tau_\mathit{PS}}
\newcommand{\TSU}{\tau_\mathit{SU}}
\newcommand{\TSUT}{\tau_{\mathit{SU}_T}}
\newcommand{\ueT}{\left(\ue^1\mathit{Pe}^1+\ue^2\mathit{Pe}^2\right)}
\newcommand{\ue}{\mathbf{u}_e}
\newcommand{\ueb}{\mathbf{u}_e^\gamma}
\newcommand{\Peb}{\mathit{Pe}^\gamma}

\newcommand{\bini}{\beta_\mathit{initial}}

\newcommand{\width}{0.45\textwidth} 

\newlength{\ckheight}
\newsavebox{\ckbox}

\let\oldequation\equation
\let\oldendequation\endequation
\let\oldalign\align
\let\oldendalign\endalign

\renewenvironment{equation}
  {\linenomathNonumbers\oldequation}
  {\oldendequation\endlinenomath}
  \renewenvironment{align}
  {\linenomathNonumbers\oldalign}
  {\oldendalign\endlinenomath}
\begin{document}

\begin{frontmatter}
\begin{abstract}
A method for density-based topology optimization of heat exchangers with two fluids is proposed. The goal of the optimization process is to maximize the heat transfer from one fluid to the other, under maximum pressure drop constraints for each of the fluid flows. A single design variable is used to describe the physical fields. The solid interface and the fluid domains are generated using an erosion-dilation based identification technique, which guarantees well-separated fluids, as well as a minimum wall thickness between them.
Under the assumption of laminar steady flow, the two fluids are modelled separately, but in the entire computational domain using the Brinkman penalization technique for ensuring negligible velocities outside of the respective fluid subdomains.
The heat transfer is modelled using the convection-diffusion equation, where the convection is driven by both fluid flows. A stabilized finite element discretization is used to solve the governing equations.
Results are presented for two different problems: a two-dimensional example illustrating and verifying the methodology; and a three-dimensional example inspired by shell-and-tube heat exchangers. The optimized designs for both cases show an improved heat transfer compared to the baseline designs. For the shell-and-tube case, the full freedom topology optimization approach is shown to yield performance improvements of up to 113\% under the same pressure drop.
\end{abstract}
\end{frontmatter}


\section{Introduction}
\label{sec:intro}
Heat exchangers are devices that serve to transfer thermal energy between two or more fluids, usually separated by solid walls to avoid mixing. They can be used for both cooling and heating applications, with some of the most well-known being in combustion engine cooling, air conditioning, power production and refrigeration.

\begin{figure*}[t]
\centering
\includegraphics[width=\textwidth]{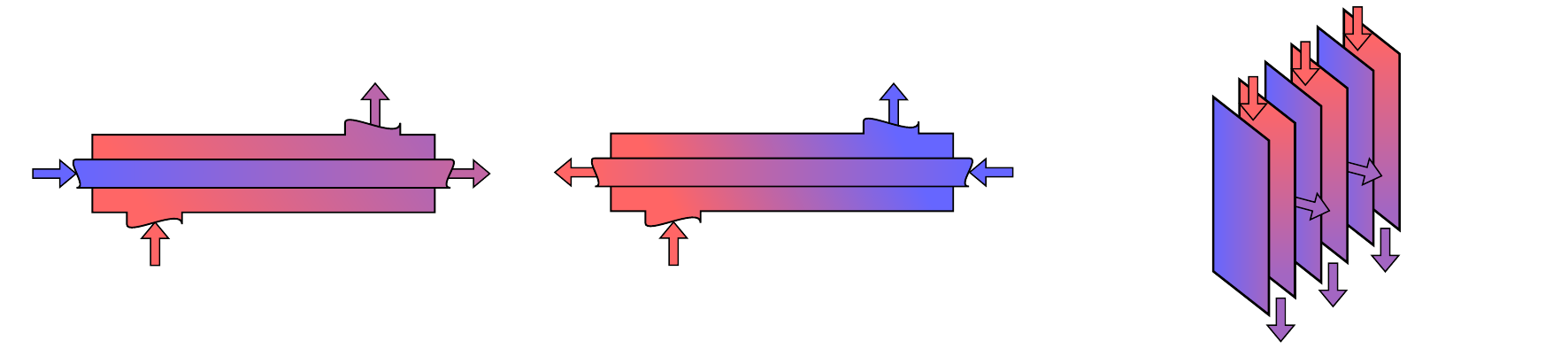}
\caption{Illustration of common heat exchanger configurations. From left: Parallel flow, counter-flow and cross-flow heat exchangers.}
\label{fig:hex_types}
\end{figure*}

Heat exchangers are widely used and their analysis is covered in most basic heat transfer courses and any good book on heat transfer, e.g. \cite{Incropera}. They are typically dimensioned and designed based on classical heat transfer theory under certain assumptions for predefined geometric layouts \cite{Awais2018}
. In recent time, the use of computational fluid dynamics (CFD) and conjugate heat transfer (CHT) simulations has become an indispensable tool for the analysis and design of complex heat exchangers \cite{AslamBhutta2012}.
However, their designs are still mainly restricted to the classical, and rather simple, geometries, such as parallel flow, counter-flow and cross-flow heat exchangers, as illustrated in Figure \ref{fig:hex_types}.
These can be assembled from standard components and manufacturing processes, e.g. punching and brazing, ensuring easy mass production at low cost.

Recent advances within additive manufacturing of conductive metals have spurred an increase in the internal geometric complexity of new heat exchanger designs \cite{Conflux,nTopology}. However, emphasis is put on maximizing the internal interacting surface area, using for instance Triply Periodic Minimal Surfaces (TPMS) \cite{nTopology}. This relies on a critical assumption based on Newton's law of cooling for the heat flux due to convection $q_{\text{conv}} = h\,A_s (T_{s} -T_{\infty})$, where $h$ is the convection heat transfer coefficient, $A_{s}$ is the area of the heat transfer surface, $T_{s}$ is the temperature of the surface and $T_{\infty}$ is the fluid reference temperature. It seems obvious to increase the surface area, $A_{s}$, to increase the heat flux. But an increase in the surface area due to increasing geometric complexity almost certainly leads to a decrease in the convection coefficient, $h$, since the fluids will flow slower due to a higher flow resistance. Of course, one can always use a more powerful pump to circumvent this, but the increased energy input must be weighted by the overall efficiency of the heat exchanger.
Therefore, this paper proposes a novel approach for the simulation-driven design optimization of pressure-drop-constrained two-fluid heat exchangers with a separating solid conductive wall using topology optimization. The approach optimizes the heat exchanger by a direct measure of the heat exchanger efficiency, based on simulations, rather than an implicit geometric quantity such as the surface area.

Topology optimization is a computational design methodology for optimizing structures. It originated in the field of solid mechanics \cite{Bendsoee1988} and has seen widespread use there \cite{Sigmund2013} over the past three decades. As detailed in the recent review paper by \citet{Alexandersen2020}, topology optimization has been applied to a wide range of flow-based problems since the first application to Stokes flow by \citet{Borrvall2003} in 2003. 
In order to apply topology optimization to heat exchanger design, it is necessary to be able to treat three-dimensional problems with high mesh resolutions. The three-dimensionality is necessary to model the complex interactions of most heat exchangers and the high mesh resolution is mainly necessary to provide a high design freedom for topology optimization by resolving small features (e.g. thin solid walls). For large scale three-dimensional flow-based problems, previous works have treated pure fluid flow \cite{Aage2008,Challis2009,Evgrafov2014,Yonekura2017,Villanueva2017,Chen2017,Dilgen2018a} and conjugate heat transfer problems for forced convection \cite{Laniewski-Wollk2016,Haertel2017,Haertel2018,Yaji2018,Dilgen2018,Pietropaoli2019} and natural convection \cite{Alexandersen2016,Alexandersen2018,Pollini2020}.

In the context of topology optimization, literature on the design of heat exchangers is very sparse. Two papers have treated guiding channels or winglets for fin-and-tube heat exchangers \cite{Sun2018,Kobayashi2019}. However, they only consider additional flow guiding features for existing heat exchanger geometries.
Only a few works consider the design of the actual heat exchanger solid surface geometry using topology optimization.
The first is the M.Sc. thesis by \citet{Papazoglou2015} investigating both a fluid tracking model and a multi-fluid model for a density-based approach.
The second is the Ph.D. thesis by \citet{Haertel2018b} coupling two-dimensional in-plane and out-of-plane flow models using an interface model similar to our approach.
The third is the paper by \citet{Tawk2019} proposing a density-based multi-fluid approach for optimizing heat exchangers with two separate fluids and a solid.
The fourth is the conference paper by \citet{Saviers2019} which, however, provides very little technical details on the applied methodology.
Finally, very recently, after the completion of the present work, a preprint was uploaded to \textit{arXiv.org} by \citet{Kobayashi2020}. The authors also use a single design variable to parametrize two fluids and a solid. The solid is represented by intermediate design variable values, whose existence is guaranteed due to filtering of the design field. However, the approach does not appear to have thickness control of the solid and is only applied to smaller computational problems.

Interface identification techniques are used to capture the transition from one physical phase to another. Such a technique was introduced by \citet{Clausen2015,clausen2017a} for topology optimization of coated structures. The method uses the spatial gradient of the design variable to identify where coating should be applied.  More recently, a modified formulation was introduced by \citet{Luo2019}, where the design field is eroded and dilated, with the intersection of these fields identifying the interface.

The paper is organized as follows: Section \ref{sec:optimization} introduces the parametrization and states the goal and constraint of the optimization problem; Section \ref{sec:phys} details the physics of the problem and the assumptions made; Section \ref{sec:FEM} provides an brief description of the finite element formulation; Section \ref{sec:topopt} presents the proposed topology optimization methodology; Section \ref{sec:res} show optimized heat exchangers for two numerical examples; and Section \ref{sec:conclusion} provides a discussion and conclusions.

\section{Parametrization}
\label{sec:optimization}
The goal of the optimization problem is to maximize the heat transferred in a heat exchanger at a given operational power, which is proportional to the pressure drop across the heat exchanger. 

\begin{figure}[ht]
    \centering
    \includegraphics[width=0.65\linewidth]{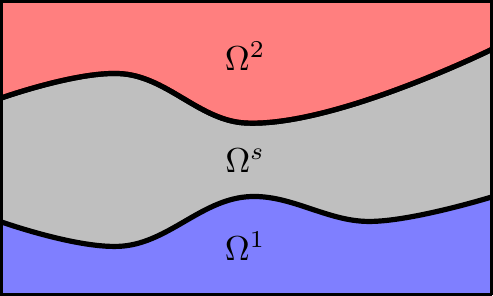}
    \caption{Sketch of the subdomains in the domain of interest. The $\Omega^1$ (blue) and $\Omega^2$ (red) domains are the domains of fluids 1 and 2, respectively. The solid domain, $\Omega^s$ (grey), separates the two fluids.}
    \label{fig:domains}
\end{figure}

The problem concerns the arrangement of a solid interface, which separates the two fluids, as illustrated in Figure \ref{fig:domains}. The fluid domains are denoted $\Omega^\gamma$, where superscript $\gamma$ is the fluid index, and are separated by a solid domain $\Omega^s$. The total computational domain is given as the union of all subdomains, $\Omega=\Omega^1\cup\Omega^2\cup\Omega^s$. In order to introduce the design representation the non-overlapping domain representation is relaxed, such that both fluids may be present in the entire domain. However, by the use of an identification technique, every point in the domain is sought to be exclusively assigned either as one of the two fluids or as solid. This means that the velocity of a fluid should be zero outside of its domain $\Omega^\gamma$.

\section{Governing equations}
\label{sec:phys}
A two fluid heat exchanger contains two mass transfer problems (one for each fluid) and one global heat transfer problem. The governing equations are derived under the assumption of steady state, constant fluid properties and incompressibility. Furthermore, heat generated by viscous dissipation is neglected. The stated assumptions result in a weak coupling between the mass and heat transfer, as only the mass transfer affects the heat transfer, but not the other way around.

In the following, the governing equations are presented in their dimensionless form. 

\subsection{Mass transfer}
\label{sec:masstrans}
The mass transfer for each of the fluid domains is obtained by solving the dimensionless Navier-Stokes equations. The equations for each mass transfer are posed in the entire domain $\Omega$ to permit topology optimization. A Brinkman friction term is used to penalize fluid flow outside the fluid subdomain \cite{Borrvall2003,Gersborg-Hansen2005,Evgrafov2005}.

For each fluid, denoted by index\footnote{Unlike for subscripts, a repeated superscript $\gamma$ does not imply summation over the index.} $\gamma$ the non-dimensional equations yield:
\begin{align}
    u_i^\gamma\diff{u_j^\gamma}{x_i}-\frac{1}{\Re^\gamma}\diff{}{x_i}\left(\diff{u_i^\gamma}{x_j} + \diff{u_j^\gamma}{x_i}\right)+\diff{P^\gamma}{x_i}&=-\alpha^\gamma(\x) u_i^\gamma\label{eq:NS}\\
    \diff{u_i^\gamma}{x_i}&=0\label{eq:masscons}
\end{align}
where $u$ is velocity, $P$ the dynamic pressure and $\alpha$ the impermeability. The Reynolds number, $\Re$, is a dimensionless parameter indicating the ratio between the inertial and the viscous forces in the flow. It is expressed as a function of a reference velocity, $U$, a length scale $L$, the fluid mass density $\rho$ and dynamic viscosity $\mu$:
\begin{equation}
    \label{eq:Re}
    \Re = \frac{UL\rho}{\mu}
\end{equation}

The impermeability, $\alpha(\x)$, is defined for each point in the domain:
\begin{equation}
    \label{eq:alpha}
    \alpha^\gamma(\x)=\left\{
    \begin{array}{ll}
        0 & {\rm if}\;\x\in\Omega^\gamma \\
        \infty & {\rm if}\;\x\notin\Omega^\gamma
    \end{array}
    \right.
\end{equation}
where it is seen that the impermeability is always $\alpha^\gamma=\infty$ outside of fluid $\gamma$. In practice, the impermeability can not be set to $\alpha=\infty$ for numerical reasons. Instead, a large value is used. For consistency, the impermeability  outside of the fluid region is related to the Darcy number \cite{Olesen2006a} and is given as:
\begin{equation}
    \label{eq:Da}
    \abmax=\frac{1}{\Re^\gamma}\frac{1}{\mathit{Da}}
\end{equation}

The mass transfer problems are subject to homogeneous Dirichlet boundary conditions on the velocity at the domain boundaries, not being in- or outlets. A homogeneous Dirichlet boundary condition is placed on the pressure at the outlet. 

\subsection{Heat transfer}
\label{sec:heattrans}
The heat transfer is described by the convection-diffusion equation. The equation is non-dimensionalised using the solid conductivity, $k^s$.  As the velocities are assumed to be $u_i^\gamma=0$ outside of their respective corresponding domain $\Omega^\gamma$, the heat transfer in the entire computational domain $\Omega$ can be expressed as:
\begin{equation}
    \label{eq:CD_ND}
    \sum_{\gamma=1}^\mathit{NF} \left(\Pes^\gamma u_i^{\gamma}\right)\diff{T}{x_i}-\diff{}{x_i}\left(\Ck(\x)\diff{T}{x_i}\right)=0
\end{equation}
where $\Pes^\gamma$ is the solid Peclet number, which relates the convective heat transfer in a fluid to the diffusive heat transfer in the solid. The conductivity ratio, $\Ck^{\gamma}$, is the fluid conductivity normalized by the solid conductivity. These parameters can be linked to the conventional Peclet number of each fluid:
\begin{align}
    \label{eq:NDparamCD}
    \Pes^\gamma&=\frac{\rho^\gamma c_p^\gamma U L}{k^s}&&,&\Ck^\gamma&=\frac{k^\gamma}{k^s} &&,& \Pe^\gamma&=\frac{\Pes^\gamma}{\Ck^\gamma}
\end{align}
The heat transfer problem is modelled by one equation for the entire domain $\Omega$, with a spatially varying coefficient $\Ck$ defined by:
\begin{equation}
    \label{eq:Ck}
    \Ck(\x)=\left\{
    \begin{array}{ll}
        \Ck^\gamma & {\rm if}\;\x\in\Omega^\gamma \\
        1 & {\rm if}\;\x\in\Omega^s
    \end{array}
    \right.
\end{equation}
The boundary conditions for the heat transfer problem consist of Dirichlet boundary conditions at the respective fluid inlets. On the rest of the domain boundary, a homogeneous Neumann boundary condition is imposed, resulting in the design domain being externally insulated.

\section{Finite element formulation}
\label{sec:FEM}
The equation system is discretized and solved using the Finite Element Method (FEM), using structured meshes with regular trilinear hexahedral elements. PSPG stabilization is employed to facilitate the use of equal-order elements. SUPG stabilization is applied to both mass transfer problems and the heat transfer problem to alleviate problems with steep gradients due to convection. 
The implementation from \cite{Alexandersen2016} is reused here without the Boussinesq approximation terms. 
As the coupling between the mass and heat transfer considered here is weak, the problems are solved sequentially. 


The two mass transfer problems are solved by finding the solution to the vector of residual equations given by:
\begin{equation}
\label{eq:NSres}
\mathbf{R}_\mathit{F\gamma}=\mathbf{M}\left(\mathbf{u}_\mathit{F\gamma},\bm{\alpha}_\gamma\right)\mathbf{u}_\mathit{F\gamma}-\mathbf{b}_\mathit{F\gamma} = \mathbf{0}
\end{equation}
The solution vector $\mathbf{u}_\mathit{F\gamma}$ contains all three velocity components and the pressure for every node. The system matrix $\mathbf{M}\left(\mathbf{u}_\mathit{F\gamma},\bm{\alpha}_\gamma\right)$ contains the viscosity, convection, Brinkman penalization, pressure coupling and velocity divergence contributions, as well as all the corresponding SUPG and PSPG stabilization terms.
The weak form and stabilization parameters are detailed in \ref{app:NSassembly}.

Similarly,
the residual equations for the heat transfer problem are defined as:
\begin{equation}
\label{eq:CDres}
\mathbf{R}_T=\mathbf{M}_T\left(\mathbf{C_k},\,\mathbf{u}_\mathit{Pe}\right)\mathbf{T}-\mathbf{b}_T = \mathbf{0}
\end{equation}
where the system is built based on the global velocity field combining both fluid flows:
\begin{equation}
\mathbf{u}_\mathit{Pe} = \Pes^1\mathbf{u}_\mathit{F1}+\Pes^2\mathbf{u}_\mathit{F2}
\end{equation}
which is possible due to the assumption of the fluid domains being well-separated at the final design.
The solution vector $\mathbf{T}$ contains the temperature in every node and the system matrix $\mathbf{M}_T\left(\mathbf{C_k},\,\mathbf{u}_\mathit{Pe}\right)$, is 
assembled from 
the thermal diffusion, convection and SUPG stabilization contributions. The weak form of these contributions, as well as the stabilization parameter, are detailed in \ref{app:CDassembly}.

\section{Topology optimization}
\label{sec:topopt}
\subsection{Optimization problem}
\label{sec:optprob}
The generic optimization problem is given as the minimization of the objective function $\Phi$, subject to $m$ constraints $g_i$. Furthermore, a nested formulation is used where the residuals of the mass- and heat transfer problems from \eqref{eq:NSres} and \eqref{eq:CDres}, are assumed zero in each iteration. The design variable $\xi$ is relaxed from discrete $\xi\in\{0,\;1\}$ to continuous $\xi\in[0,\;1]$ and represented by $n$ elementwise constant scalars. 
\begin{equation}
\begin{split}
    \min_{\xi\in\left[0,\;1\right]^n}\;&\Phi(\xi,\;u^1,\;u^2,\;T)\\
    \mathrm{s.t.}\;&\mathbf{R}_{F\gamma}=0,\;\gamma=\{1;\,2\}\\
    &\mathbf{R}_T=0\\
    &g_i\leq0,\;i=1\dots m
\end{split}
\end{equation}

The objective function for a heat exchanger is to maximize the thermal energy transferred from the hot to the cold fluid. This can be expressed as minimizing the difference between the enthalpy flowing out at the cooled and coolant fluid outlets\footnote{Figure \ref{fig:parchan_domain} shows the formal definition of the inlet and outlet regions.}:
\begin{equation}
    \label{eq:entalpy_obj_hx}
    \Phi = \frac{1}{\int_{\Gamma_{F2}} dA}\left(\Pes^1\int_{\Gamma_{F1}} n_iu^1_iTdA-\Pes^2\int_{\Gamma_{F2}} n_iu^2_iTdA\right)
\end{equation}
This objective function has the advantage of being defined on both fluid outlets, which it is beneficial for the computation of the sensitivities, as both mass transfer adjoint problems will have a source term on their respective outlets. 
However, it can be difficult to associate physical meaning to it. Therefore, for comparison purposes, we introduce $\phih$, which is an expression of the heat transferred to the coolant, normalized by the outlet area:
\begin{equation}
    \label{eq:phih}
    \phih= \frac{\Pes^2}{\int_{\Gamma_{F2}} dA}\int_{\Gamma_{F2}} n_iu^2_iTdA
\end{equation}
where a higher value is preferred since it reflect the amount of heat transferred to the coolant from the other fluid.

In order to regularize the geometry, and impose restrictions on the pumping power, the pressure drop on each fluid phase is controlled. This is done by placing a pressure drop constraint on each of the fluids:
\begin{equation}
    \label{eq:dp_gamma}
    g_\gamma = \frac{1}{\Delta P_{max}^\gamma \int_{\Gamma_{in,\;F\gamma}} dA}\int_{\Gamma_{in,\;F\gamma}}P^\gamma dA-1
\end{equation}
where $\dpmax^{\gamma}$ is the maximal admissible pressure drop on fluid $\gamma$.

The optimization problem is solved using the Method of Moving Asymptotes (MMA) \cite{SVANBERG1987}, implemented in PETSc \cite{Aage2015} using external move limits of $0.2$.

\subsection{Filtering and projecting}
\label{sec:filterproject}
In topology optimization, filtering techniques are used to prevent checkerboards and other unwanted effects from appearing in the obtained designs \cite{Sigmund1998}. The PDE filter \cite{Lazarov2011,Aage2015} is here used to obtain a filtered design field $\hat{\xi}$:
\begin{align}
    \label{eq:topopt_PDEfilter}
    -R^2\nabla^2\hat{\xi}+\hat{\xi}&=\xi\\
    \label{eq:topopt_PDErad}
    R &= \frac{r}{2\sqrt{3}}
\end{align}
where $r$ is the physical filter radius.

A smooth Heaviside projection \cite{Wang2011} with threshold $\eta$ and sharpness $\beta$ is applied to the filtered variable, which leads to the projected field $\tilde{\xi}$:
\begin{equation}
    \label{eq:topopt_heavi}
    \tilde{\xi}(\hat{\xi},\beta,\eta) =\frac{\tanh(\beta\eta)+\tanh(\beta(\hat{\xi}-\eta))}{\tanh(\beta\eta)+\tanh(\beta(1-\eta))}
\end{equation}

In combination with the filter, the Heaviside projection, can be used to enforce a minimum length scale on the obtained structures \cite{Wang2011} if multiple design realizations are considered.

\subsection{Modelling solid interfaces}
\label{sec:erodila}
A single design variable, $\xi$, is used to model three physical phases: two fluids and one solid. In order to ensure strict separation between the two fluid phases, it is important that there always exists a solid wall between them. A method for the introduction of a third phase between two already existing ones, has been introduced for minimization of elastic compliance of coated structures \cite{Clausen2015,clausen2017a,Luo2019}. In its original application, the purpose of this method is to introduce a coating of a specific thickness between the void and the solid. In the present case, solid is placed between two fluids and allows for rigorous control of the interface thickness.

\begin{figure}[tb]
    \centering
    \includegraphics[width=\linewidth]{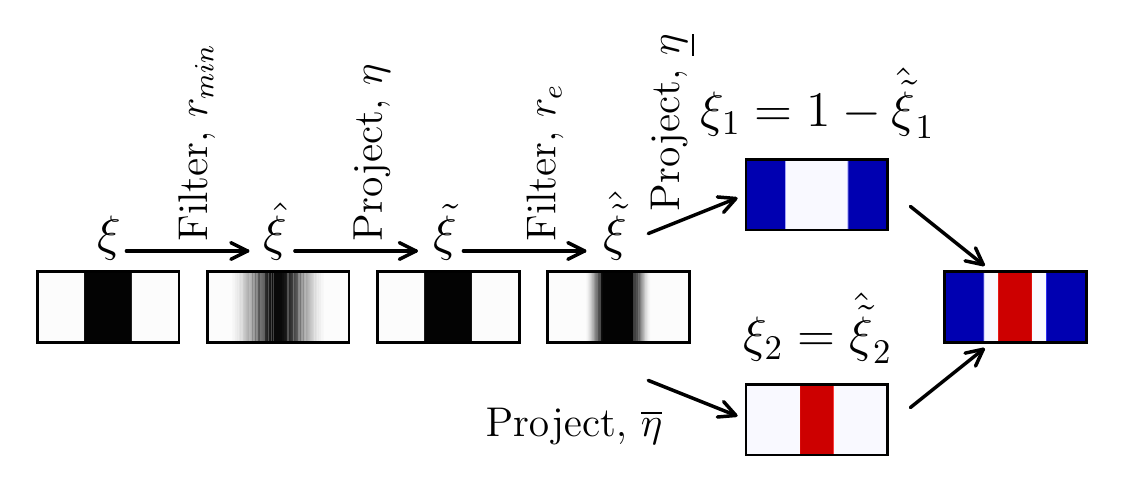}
    \caption{Overview of the different variables used in the erosion-dilation process. The resulting variables $\xiI$ and $\xiII$ are used to define the domains of the two fluids (red and blue) and of the solid interfaces (white).}
    \label{fig:ero_vars}
\end{figure}

As seen in Figure \ref{fig:ero_vars}, the process consists of a filtering operation with filter radius $r_\mathit{min}$ and a projection using the threshold $\eta=0.5$. This is done to ensure a well-defined interface. The obtained variable $\xit$ is refiltered using the erosion radius $r_e$, leading to the re-filtered variable $\xip$. Finally, the re-filtered variable $\xip$ is projected at a low and a high threshold, $\left\{\underline{\eta},\;\overline{\eta}\right\}=0.5\pm\Delta\eta$, yielding intermediate variables $\hat{\Tilde{\xi}}_{1}$ and  $\hat{\Tilde{\xi}}_{2}$, respectively. The eroded and dilated variables, indicate which physical phase is applicable in each element (where the dilated variable, $\xihtI$ is mapped for consistency):
\begin{equation}
    \label{eq:topopt_whichfluid}
    \left\{
    \begin{array}{ll}
        \text{Fluid 1:}&\;\xiI=1-\xihtI=1\\
        \text{Fluid 2:}&\;\xiII=\xihtII=1\\
        \text{Solid:}&\;\xiI=\xiII=0
    \end{array}\right.
\end{equation}

Two advantages of the erosion-dilation technique should be mentioned here, one is that a separating solid phase is guaranteed between the two fluids and the other is that the thickness of this solid phase can be controlled. As shown by \cite{Luo2019}, a wall thickness $w_e$ is obtained by using the high and low threshold values and erosion filter radius:
\begin{equation}
    \label{eq:topopt_re_we}
    \Delta\eta=0.45 \qquad   r_e\approx0.75w_e
\end{equation}
It is noted that the first filtering operation should have a filter radius larger than the wall thickness:
\begin{equation}
    \label{eq:topopt_rminerosion}
    r_\mathit{min}>w_e
\end{equation}

\subsection{Interpolation functions}
\label{sec:interpfunc}
The relaxation of the optimization problem introduces the need for interpolation functions. Two parameters, the impermeabilities of both fluids $\alpha^\gamma$ and the conductivity ratio $C_k$ are interpolated from the variables obtained by the erosion-dilation process seen in Figure \ref{fig:ero_vars}. 

The impermeabilities $\alpha^\gamma$ are interpolated from the corresponding variable from the erosion-dilation process $\xi_\gamma$. The interpolation is done using RAMP \cite{Borrvall2003,Stolpe2001,alexandersen2013a}:
\begin{equation}
    \label{eq:topopt_RAMPalpha}
    \alpha^\gamma(\xi_\gamma)=\abmax\frac{1-\xi_\gamma}{1+\xi_\gamma q_\alpha}
\end{equation}
where the upper limit $\abmax$ is the impermeability to be applied where the fluid $\gamma$ is not present, as discussed in Section \ref{sec:masstrans}. The parameter $q_\alpha$ indicates the curvature of the function, which is linear when $q_\alpha=0$.

The conductivity ratio $\Ck$ for the heat transfer problem is interpolated from both $\xiI$ and $\xiII$. The interpolation has three bounds, such that Equation (\ref{eq:Ck}) is fulfilled. This is done by introducing the following interpolation function inspired by SIMP \cite{Bendsoee1999}:
\begin{equation}
    \label{eq:topopt_SIMPCk}
    C_k(\xiI,\;\xiII) = \left(1-\xiI-\xiII\right)^{p_k}+C_k^1\xiI+C_k^2\xiII
\end{equation} 
It is seen that the introduced interpolation function for $\Ck$ has a penalization power $p_k$ on the solid phase, but not on the two terms corresponding to the fluid phases. Numerical studies concluded that this formulation circumvents high relative conductivities, when both $\xiI$ and $\xiII$ have intermediate values. In Figure \ref{fig:Ckinterp}, the interpolation function is shown for different penalization powers $p_k$. 

\sbox\ckbox{%
  \resizebox{\dimexpr\textwidth}{!}{%
    \includegraphics[height=3cm]{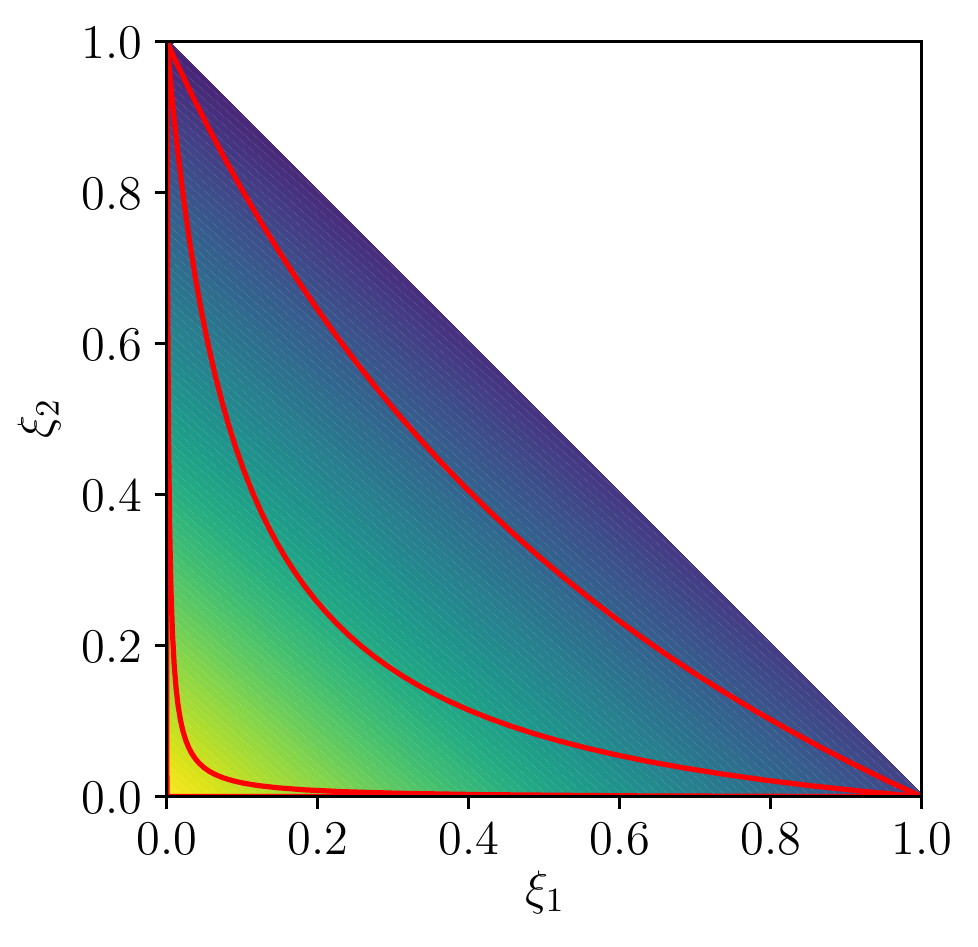}%
    ~
    \includegraphics[height=3cm]{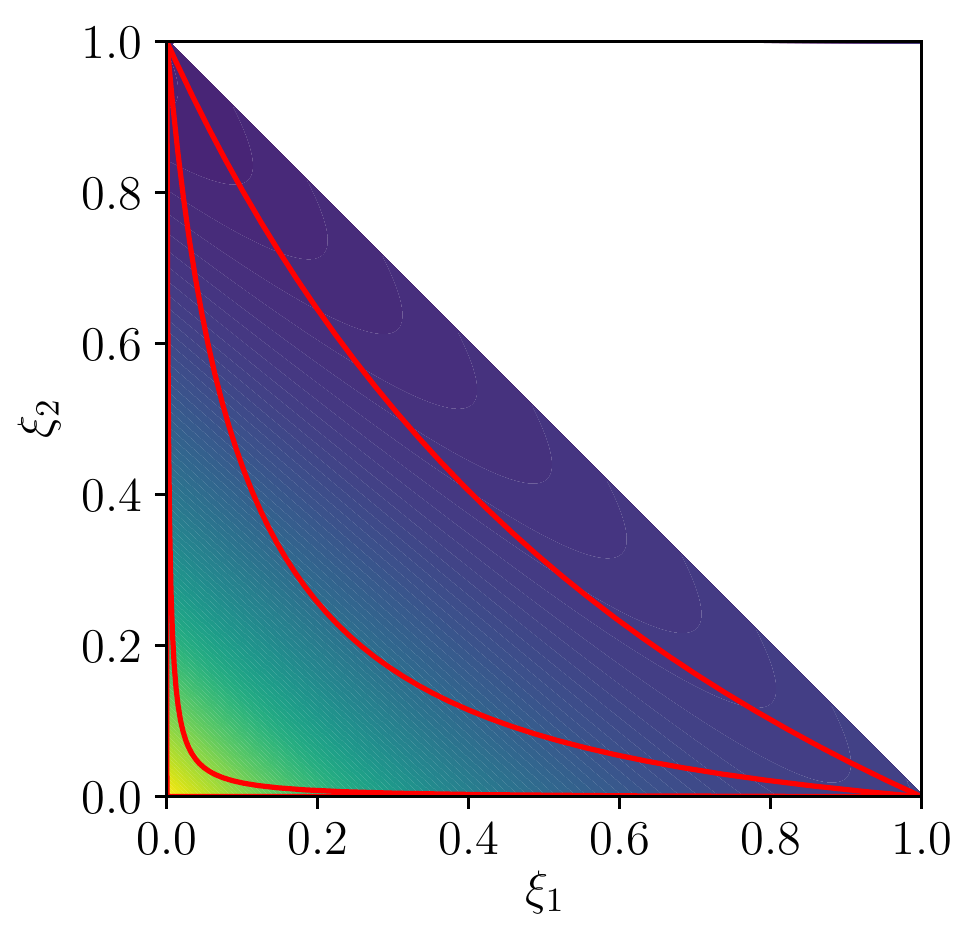}%
    ~
    \includegraphics[height=3cm]{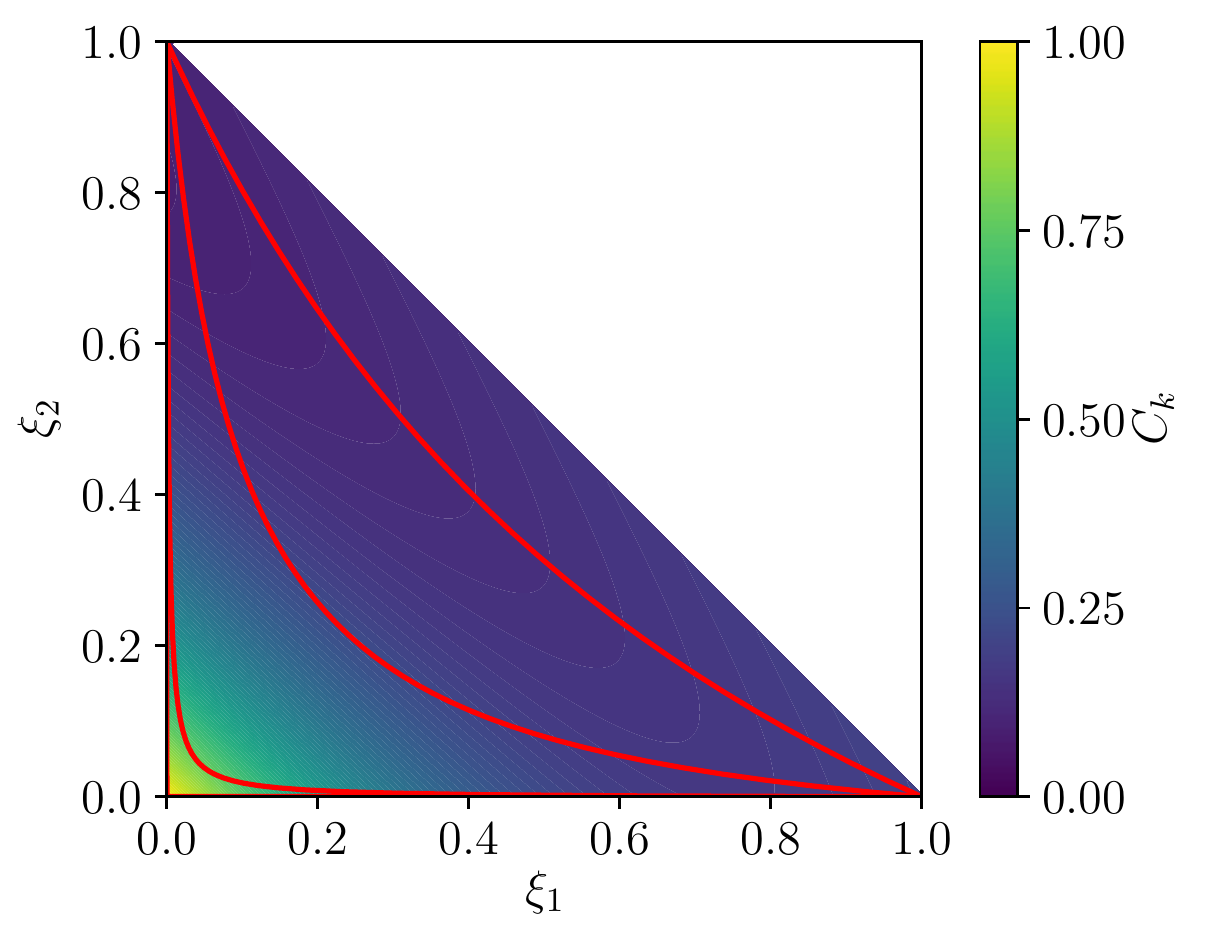}%
  }%
}
\setlength{\ckheight}{\ht\ckbox}

\begin{figure*}[htb]
    \centering
    \subcaptionbox{$p_k=1$}{\includegraphics[height=\ckheight]{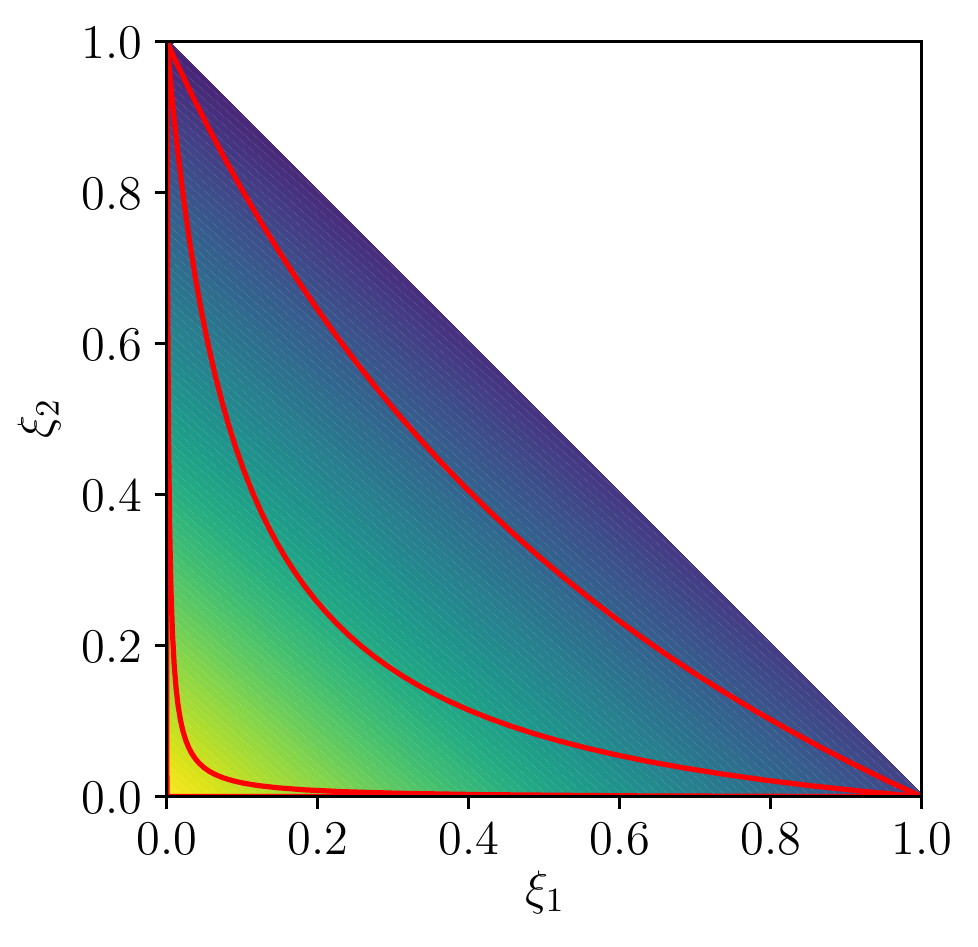}}
    ~
    \subcaptionbox{$p_k=2$}{\includegraphics[height=\ckheight]{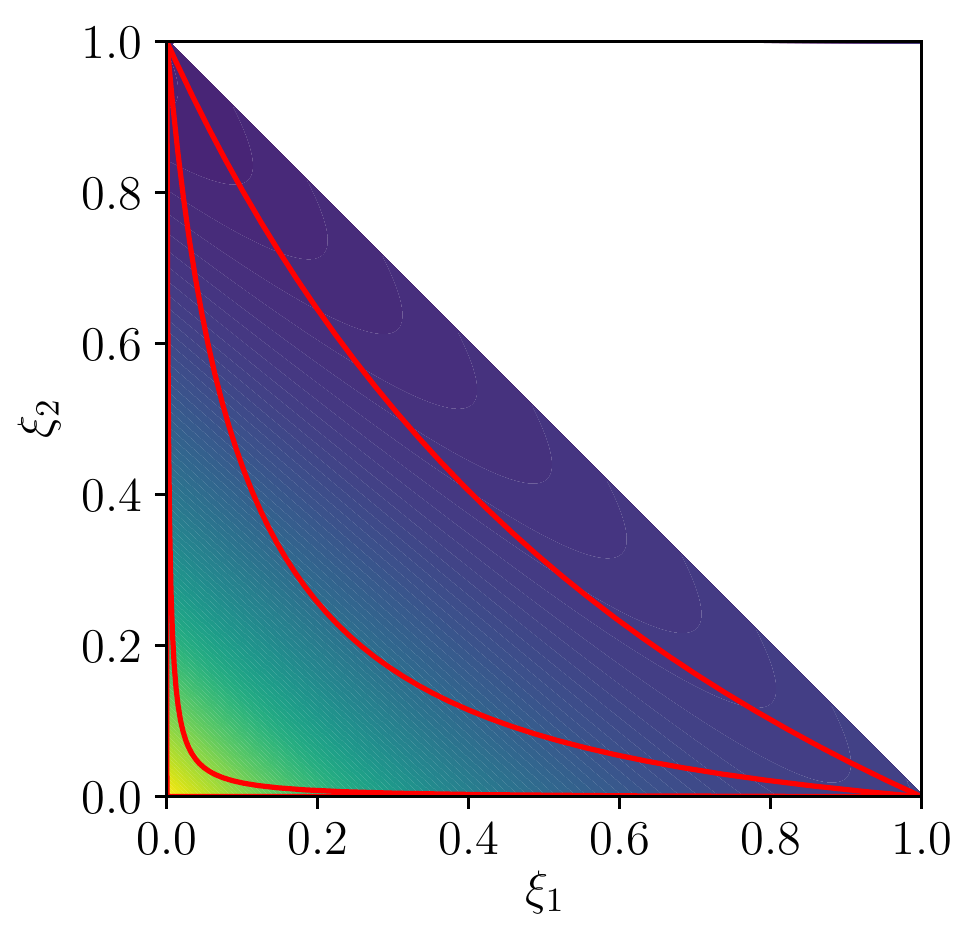}}
    ~
    \subcaptionbox{$p_k=3$}{\includegraphics[height=\ckheight]{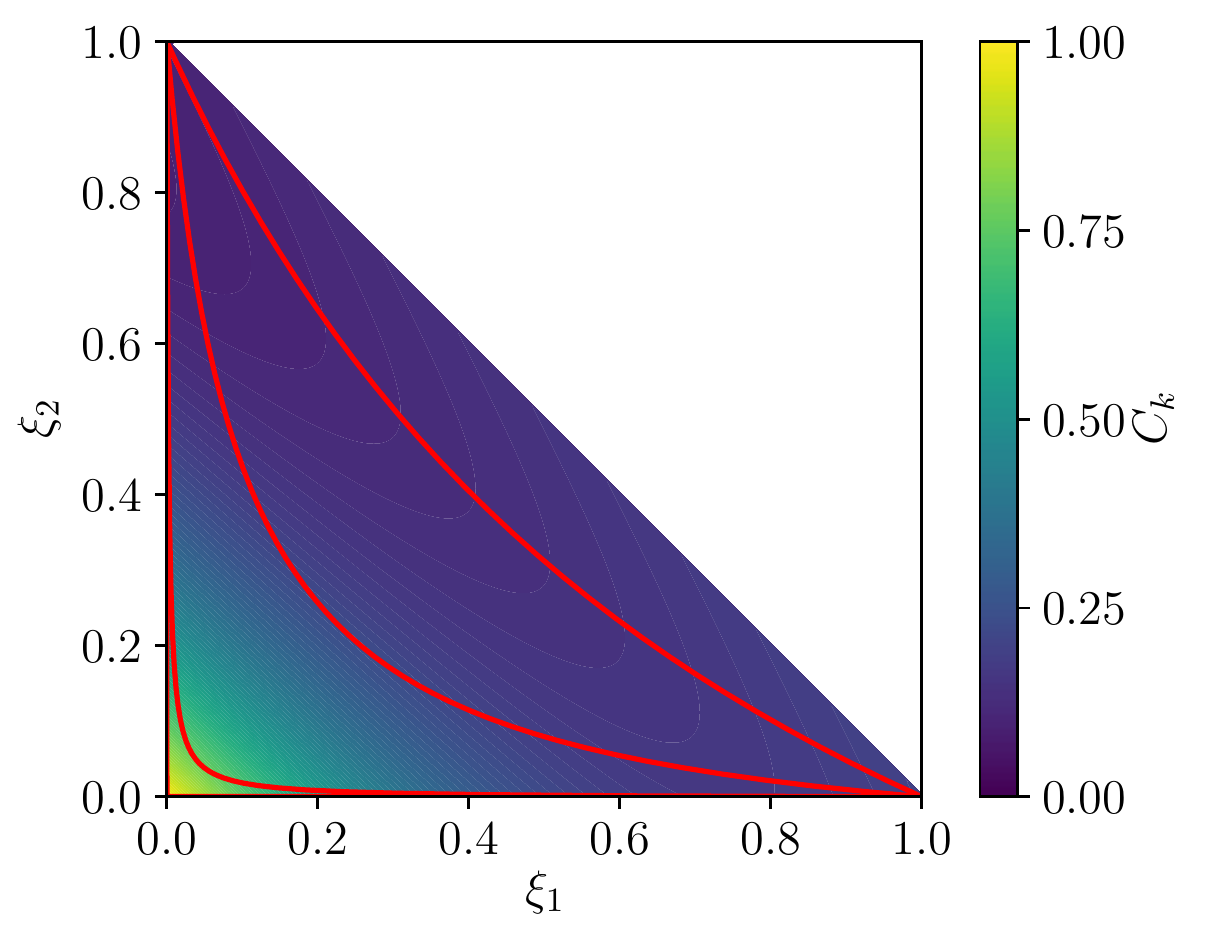}}
    \caption{Interpolation function for the relative conductivity, with $C_k^1=0.2$ and $C_k^2=0.1$. The red lines show the $\xiI$ and $\xiII$ combinations obtained with the identification method for $\beta=\left\{1,\;2,\;4\right\}$. Note that the northeastern part is white, as the corresponding combinations between $\xiI$ and $\xiII$ are not feasible.}
    \label{fig:Ckinterp}
\end{figure*}

\subsection{Sensitivity analysis}
\label{sec:optsens}
The sensitivities of the objective and constraint functions are determined using the adjoint method. The method consists of setting up a Lagrangian function, where the residuals of the FEM problems are multiplied with the Lagrangian multipliers $\bm{\lambda}_i$ (also known as the adjoint variables):
\begin{equation}
\label{eq:lagrangian}
    \mathcal{L} = \Phi + \bm{\lambda}_{F1}^\intercal\mathbf{R}_{F1} + \bm{\lambda}_{F2}^\intercal\mathbf{R}_{F2}+\bm{\lambda}_T^\intercal\mathbf{R}_T
\end{equation}

The derivative of the Lagrangian function is derived and rewritten using the chainrule, as seen in \ref{app:adjoint}, which results in the following sensitivity expression:
\begin{equation}
\label{eq:sens_2f}
    \ddiff{\Phi}{\xi}=\diff{\Phi}{\xi}+\bm{\lambda}_T^\intercal\diff{\mathbf{R}_T}{\xi}+\bm{\lambda}_{F1}^\intercal\diff{\mathbf{R}_{F1}}{\xi}+\bm{\lambda}_{F2}^\intercal\diff{\mathbf{R}_{F2}}{\xi}
\end{equation}
where the Lagrangian multipliers are found by solving the three weakly coupled adjoint problems:
\begin{align}
    \left(\diff{\mathbf{R}_T}{T}\right)^\intercal\bm{\lambda}_T&=\left(-\diff{\Phi}{T}\right)^\intercal\label{eq:sens_lT}\\
    \left(\diff{\mathbf{R}_{F1}}{\mathbf{u_1}}\right)^\intercal\bm{\lambda}_{F1}&=-\left[\left(\diff{\Phi}{\mathbf{u_1}}\right)^\intercal+\left(\diff{\mathbf{R}_T}{\mathbf{u_1}}\right)^\intercal\bm{\lambda}_T\right]\label{eq:sens_lF1}\\
    \left(\diff{\mathbf{R}_{F2}}{\mathbf{u_2}}\right)^\intercal\bm{\lambda}_{F2}&=-\left[\left(\diff{\Phi}{\mathbf{u_2}}\right)^\intercal+\left(\diff{\mathbf{R}_T}{\mathbf{u_2}}\right)^\intercal\bm{\lambda}_T\right]\label{eq:sens_lF2}
\end{align}
The transposed tangential system matrices used in the adjoint problem are, as in the physical problem, adjusted for the imposed Dirichlet boundary conditions on the physical problem. However, it should be noted, that all Dirichlet boundary conditions in the adjoint problems are homogeneous.

\subsection{Continuation of parameters}
\label{sec:cont}
The values of the projection sharpness, $\beta$, as well as the parameters of the interpolation functions, $q_\alpha$ and $p_k$ in \eqref{eq:topopt_RAMPalpha} and  \eqref{eq:topopt_SIMPCk}, respectively, are to be set in order to obtain physical interpolation schemes. The initial parameters are chosen to give the optimizer a lot of freedom in the beginning, and modified to obtain sharper interfaces and a more accurate physical modelling as optimization progresses.

For two- and three-dimensional problems, a continuation step is applied every $40^\text{th}$ and $20^\text{th}$ design iteration, respectively, if the constraints have been met in the previous 3 iterations. Numerical studies showed that taking a single relaxation step on $q_\alpha$ performed well. When increasing the projection sharpens, the $\beta$ value is doubled. Finally, the penalization power for the relative conductivity interpolation, $p_k$, is increased simultaneously with the projection sharpness. The continuation scheme is shown in detail in Table \ref{tab:continuation}. 
\begin{table}[htb]
    \centering
    \caption{Projection sharpness and interpolation parameters as function of continuation step. Step 1 is the initial setting.}
    \label{tab:continuation}
    \begin{tabular}{l||rrrrrrr}
        Step        & $1$    & $2$    & $3$    & $4$    & $5$    & $6$    & $7$ \\\hline\hline
        $\beta$     & $1$    & $2$    & $2$    & $4$    & $8$    & $16$   & $32$\\
        $q_\alpha$ & $10^4$ & $10^4$ & $10^3$ & $10^3$ & $10^3$ & $10^3$ & $10^3$\\
        $p_k$       & $1$    & $1$    & $1$    & $2$    & $2$    & $3$    & $3$
    \end{tabular}
\end{table}
When starting with a higher projection sharpness than $\beta=1$, the scheme seen in Table \ref{tab:continuation} is used, replacing the $\beta$ value in step 1 and omitting the obsolete continuation steps.

\section{Results}
\label{sec:res}
First, a simple optimization problem in two dimensions is considered. This relatively simple problem allows for a demonstration and verification of the methodology and the design representation by one design variable. Thereafter, a more complex three-dimensional problem is considered, where the full potential of the method is demonstrated.

\subsection{Two-dimensional counter-flow heat exchanger}
\subsubsection{Problem definition}
\label{sec:probdefparchan}
\begin{figure*}[tbp]
    \centering
    \includegraphics[width=0.7\linewidth]{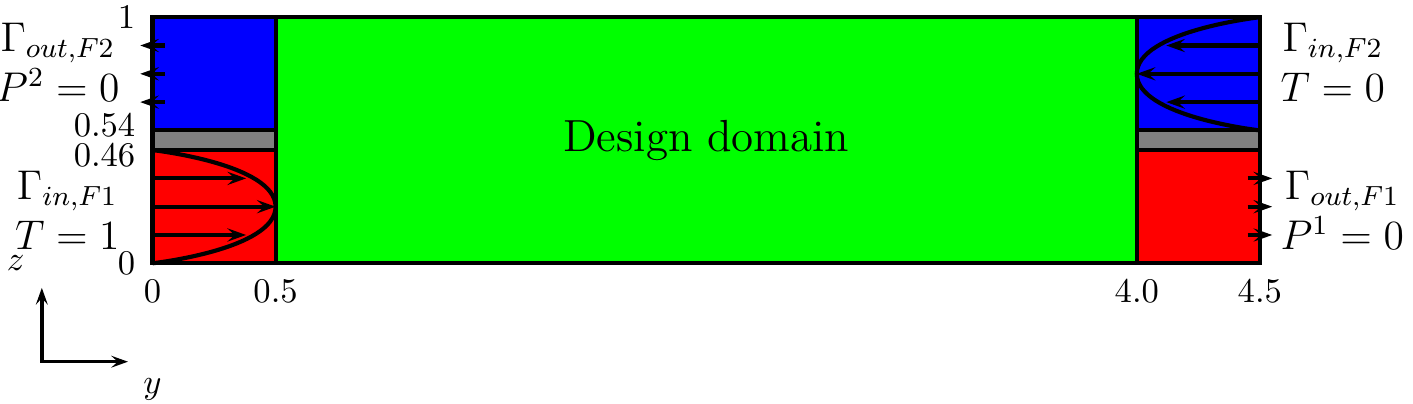}
    \caption{Sketch of the design domain for the two dimensional counter-flow heat exchanger. Green indicates the actual design domain, passive domains of the coolant- and cooled fluids are shown in blue and red, respectively. Grey indicates passive solid domains. The reference length is the height of the domain, $L=1$ and the reference velocity, the velocity at the inlet center, $U^\gamma=1$.}
    \label{fig:parchan_domain}
\end{figure*}

The first case considered is a two-dimensional counter-flow heat exchanger. The setup is seen in Figure \ref{fig:parchan_domain} and consists of a hot fluid inlet in the lower-left part of the domain, with the corresponding outlet on the opposite lower-right side of the domain. A cold fluid inlet is located at the upper-right side of the domain, with the corresponding outlet at the opposite upper-left side. The inlets have parabolic velocity profiles assuming fully-developed laminar flow. Homogeneous Dirichlet boundary conditions are applied on the pressures at the respective outlets. Furthermore, a straight out velocity boundary condition is placed on the velocities at the outlets.
The areas near the in- and outlets are excluded from the optimization to ensure undisturbed in- and outflow.

\begin{figure}
    \centering
    \begin{subfigure}{\linewidth}
    \includegraphics[width=\linewidth]{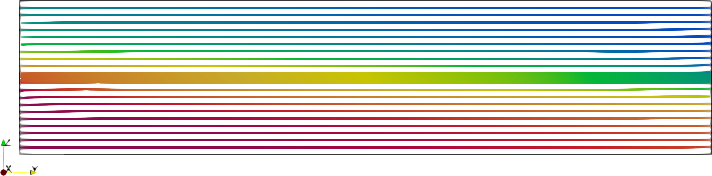}
    \end{subfigure}
    \\
    \begin{subfigure}{0.5\linewidth}
    \includegraphics[width=\linewidth]{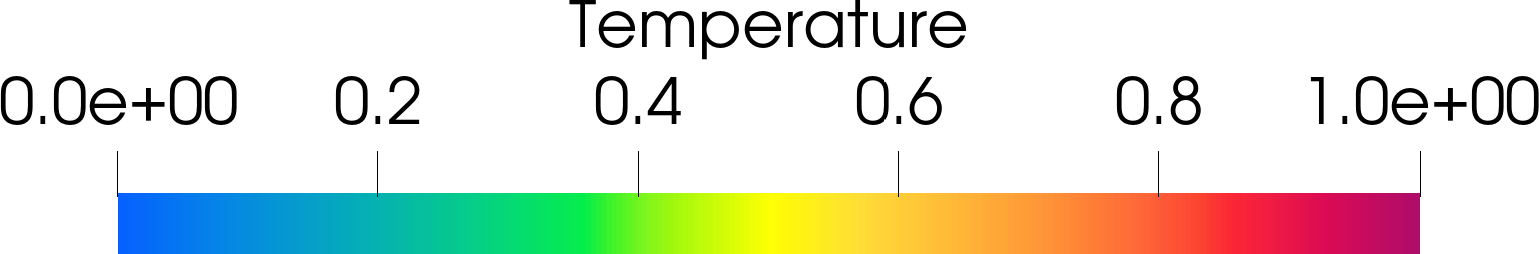}
    \end{subfigure}
    \caption{Baseline design of the two-dimensional counter-flow example. The two fluid phases consist of straight channels separated by a wall. Structure and streamlines are colored by temperature.}
    \label{fig:parchanbase}
\end{figure}

The problem presented here is optimized for computational Reynolds number $\Re_\mathit{comp}^\gamma=100$, which leads to an actual Reynolds number $\Re^\gamma=46$, for both fluids, when adjusted for their respective inlet sizes \cite{Alexandersen2016}. The reference velocity, $U$, is the maximum velocity, located at the center of the intlets. The computational Peclet number is $\Pe_\mathit{comp}^\gamma=100$, as  for the Reynolds numbers, this leads to the actual Peclet number $\Pe^\gamma=46$.
\begin{table}[tbp]
    \centering
    \caption{Computational parameters of heat exchanger}
    \label{tab:2D_param}
    \begin{tabular}{l||rr}
        Property                                & Cooled      &  Coolant\\\hline\hline
        Reynolds number $\scriptstyle\Re[-]$       & $100$        & $100$\\
        Solid Peclet number $\scriptstyle\Pes[-]$         & $4$        & $4$\\
        Conductivity ratio $\scriptstyle\Ck[-]$ & $0.04$        & $0.04$\\
        Impermeability $\scriptstyle\amax[-]$ & $10^4$ & $10^4$
    \end{tabular}
\end{table}
The computational non-dimensional parameters are seen in Table \ref{tab:2D_param}. The optimization process is cut-off after a maximum of 500 design iterations.

The optimization problem is run with different maximum pressure drop constraints, which are obtained from the empty straight-channel design, as seen in Figure \ref{fig:parchanbase}.
The objective function is compared to the one from the same baseline design:
\begin{equation}
    \label{eq:parchan_base}
    \Delta P^\gamma_\mathit{base}= 1.57\qquad\Phi_\mathit{coolant,\,base} = 0.856
\end{equation}

\subsubsection{Results}
\label{sec:res1varparchan}
The problem is optimized for a range of values of admissible pressure drops~$\dpmax$. It should be noted that the pressure drop in the channels cannot be smaller than the one from the baseline design, where the two channels are straight and parallel in the entire domain.
The domain consists of $324\times72\times 1$ cubic elements. The filter radii are set, such that the wall is $w_e=0.075$ thick, corresponding to $\approx5$ elements. Furthermore, the filter radius has been set to $r_\mathit{min} = 0.08$, as discussed in Section \ref{sec:erodila}.

Two initial designs are considered in this example. The first one consists of the two parallel channels. In the second initial design, the entire design domain has been set to an intermediate design variable value, $\xi=0.5$, violating the strict separation of fluids.

\begin{figure*}[tbp]
    \centering
    \begin{subfigure}{\width}
    \includegraphics[width=\linewidth]{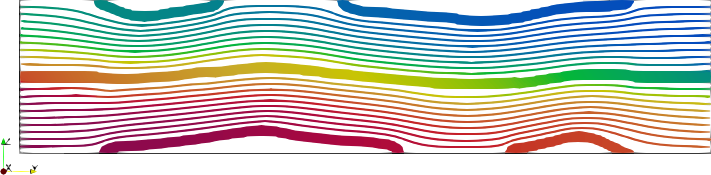}
    \caption{\dpf{2.0}, \obj{1.024}}
    \label{fig:1varparchan_2}
    \end{subfigure}%
    ~
    \begin{subfigure}{\width}
    \includegraphics[width=\linewidth]{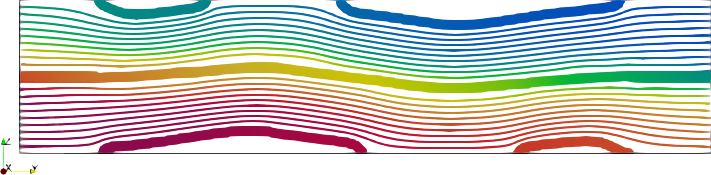}
    \caption{\dpf{2.0}, \obj{1.025}}
    \label{fig:1varpar05_2}
    \end{subfigure}
    \\
    \begin{subfigure}{\width}
    \includegraphics[width=\linewidth]{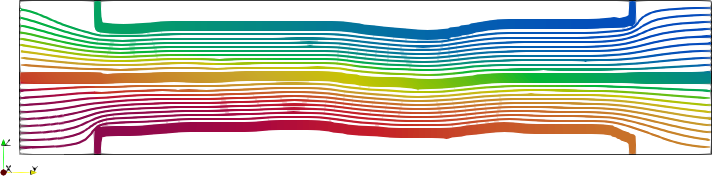}
    \caption{\dpf{5.0}, \obj{1.212}}
    \label{fig:1varparchan_8}
    \end{subfigure}%
    ~
    \begin{subfigure}{\width}
    \includegraphics[width=\linewidth]{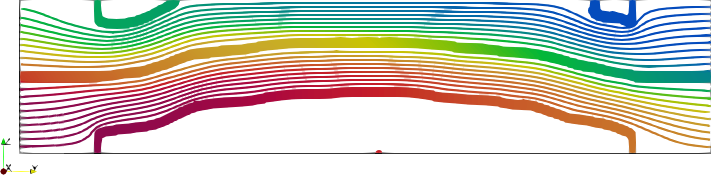}
    \caption{\dpf{5.0}, \obj{1.224}}
    \label{fig:1varpar05_8}
    \end{subfigure}
    \\
    \begin{subfigure}{\width}
    \includegraphics[width=\linewidth]{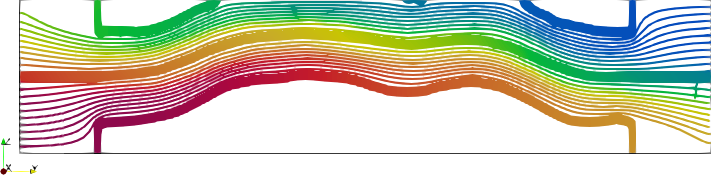}
    \caption{\dpf{10.0}, \obj{1.395}}
    \label{fig:1varparchan_18}
    \end{subfigure}%
    ~
    \begin{subfigure}{\width}
    \includegraphics[width=\linewidth]{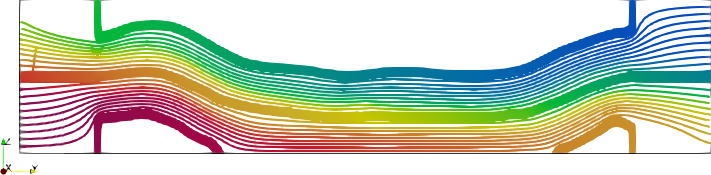}
    \caption{\dpf{10.0}, \obj{1.390}}
    \label{fig:1varpar05_18}
    \end{subfigure}\\
    \begin{subfigure}{0.3\textwidth}
    \includegraphics[width=\linewidth]{colbar_temp.png}
    \end{subfigure}
    \caption{Optimized structures and corresponding streamlines, colored by temperature. The structures are optimized with two different initial designs being, (a, c, e) the parallel channels and (b, d, f) $\xi=0.5$. The designs are optimized for different admissible pressure drops, (a, b) \dpf{2.0}, (c, d) \dpf{5.0} and (e, f) \dpf{10.0}. Structures are thresholded at $\aI>0.1\amI\land\aII>0.1\amII$.}
    \label{fig:parchan_struct}
\end{figure*}

A selection of the optimized structures, obtained at different admissible pressure drops and using the two initial designs are seen in Figure \ref{fig:parchan_struct}.
With both initial designs, the channels get narrower, as the admissible pressure drop is increased. The narrower channels increase the flow speed of both fluids, which increases the heat transfer coefficient. Furthermore, in some cases, it is seen that the channels bend towards the domain boundary. A reason for this might be that a solid wall, with a high conductivity, can be avoided along the fluid. This wall might otherwise act as an unwanted regenerator.

\begin{figure}[htb]
    \centering
    \includegraphics[width=\linewidth]{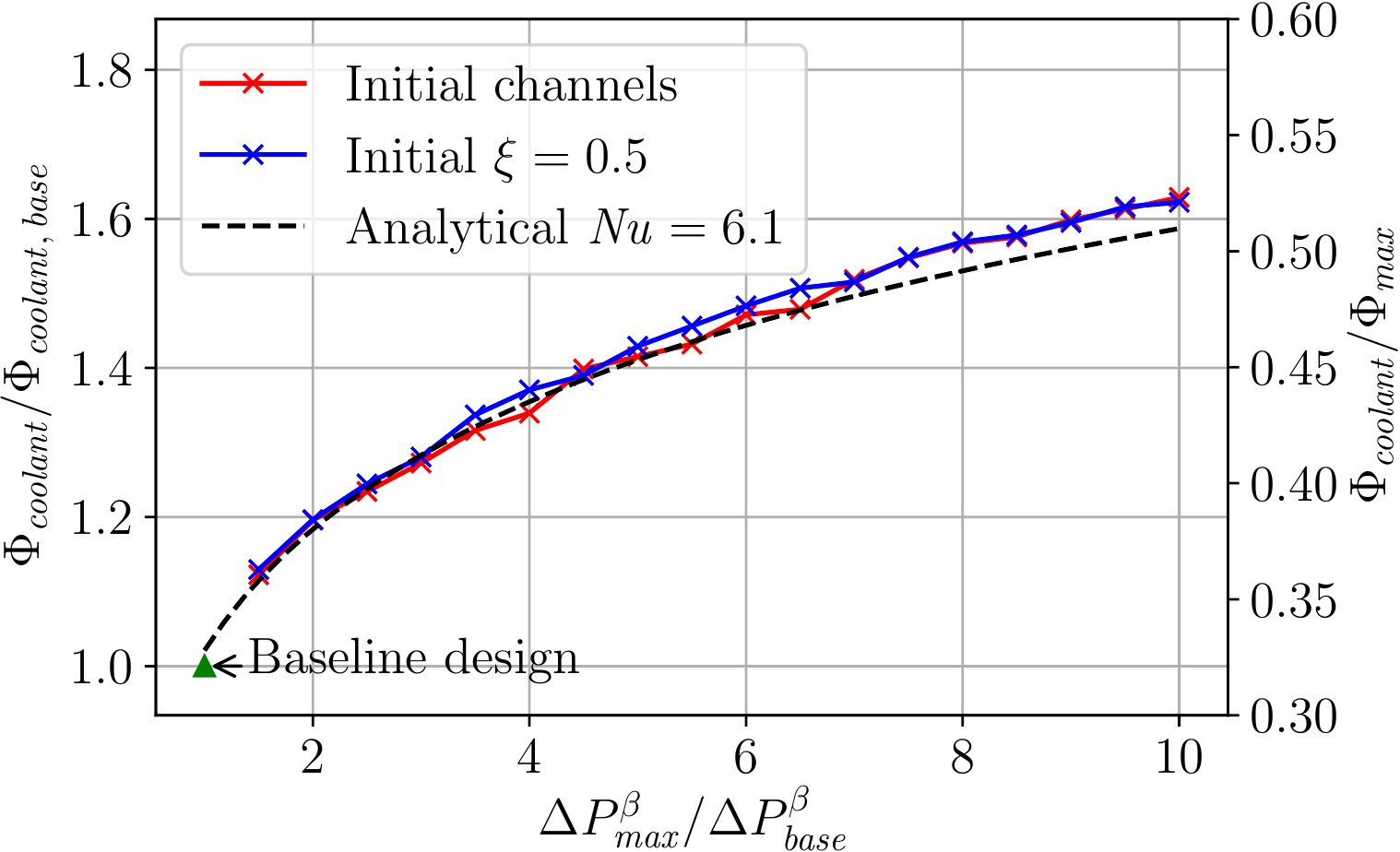}
    \caption{Improvement and fit of the transferred heat as a function of the admissible pressure drop in the fluids.}
    \label{fig:1varparchanimprov}
\end{figure}
In Figure \ref{fig:parchan_struct}, a higher heat transfer is also observed for the designs optimized with a higher admissible pressure loss. In Figure \ref{fig:1varparchanimprov}, the heat transfer improvements (relative to the baseline design and to a theoretical limit) are compared for different admissible pressure drops (relative to the baseline design). The monitored improvement can be compared to a function of the admissible pressure drop $\dpmax$. This is done by noting the efficiency $\epsilon$ as a function of the Number of Transfer Units ($\mathit{NTU}$) \cite{Thulukkanam2013}:
\begin{align}
\label{eq:NTUeps}
\epsilon&=\frac{\phih}{\Phi_\mathit{max}}=\frac{\mathit{NTU}}{1+\mathit{NTU}}\\
\mathit{NTU}&=\frac{\mathit{UA}}{c_p\dot{m}}\label{eq:NTU}
\end{align}
where $\mathit{UA}$ is the overall heat transfer coefficient, which is dependent of the heat transfer coefficients $h_{F\gamma}$ of the two fluids:
\begin{align}
\mathit{UA}&=\left(\frac{L_w}{k_wA}+\frac{1}{h_\mathit{F1}A}+\frac{1}{h_\mathit{F2}A}\right)^{-1}\label{eq:UA}\\
h_{F\gamma}&=\frac{\mathit{Nu}_\gamma k_{F\gamma}}{D_h}\label{eq:hf}
\end{align}
where $\mathit{Nu}_\gamma$ is the Nusselt number and $k_\mathit{F\gamma}$ the conductivity of the respective fluid. The Nusselt number for both fluids is found to be $\mathit{Nu}=6.1$ for this counter-flow heat exchanger problem. This was done by performing a numerical analysis of the baseline model in COMSOL. The hydraulic diameter is twice the height of the infinitely wide channel, $D_h=2w$, where $w$ is the channel height. Assuming all the pressure loss goes to making the channels narrower and that the flow is always fully developed, the width of the channel can be computed from integrating the volume flux of the Poiseuille flow:
\begin{equation}
\label{eq:volflux}
\begin{split}
V&=\int_0^w -\frac{1}{2\mu}\frac{dP}{dx}\left(yw-y^2\right)dz\\
\Rightarrow w&=w_\mathit{base}\sqrt[3]{\frac{\dpbase}{\Delta P}}
\end{split}
\end{equation}
Combining Equations (\ref{eq:NTUeps}-\ref{eq:volflux}), an analytical heat transfer enhancement for straight pipes with varying channel width $w$ can be computed and compared to the optimized designs. This is done assuming that the two fluids behave equally and that the allowable pressure drop is used exclusively to make the channels thinner on the entire length of the domain. 

From Figure \ref{fig:1varparchanimprov}, it is seen that there is a very nice correspondence between the performance of the optimized designs and the theoretical prediction. At larger admissible pressure drops, the optimized designs outperform the theoretical prediction marginally. This confirms that, at least for this example, it is possible to obtain close to optimal designs using the presented methodology. The higher heat transfer by the designs obtained using the present methodology is due to the larger design freedom, which, for instance, makes the channels slightly curved. The slight advantage of these features can be seen from the two designs optimized for \dpf{5}, where the design with the curved feature, Figure \ref{fig:1varpar05_8}, slightly outperforms ($\sim 1\%$) the designs, where the channels have been narrowed, but stayed parallel, Figure \ref{fig:1varparchan_8}.

\begin{figure*}[htb]
\centering
    \begin{subfigure}{\width}
    \includegraphics[width=\linewidth]{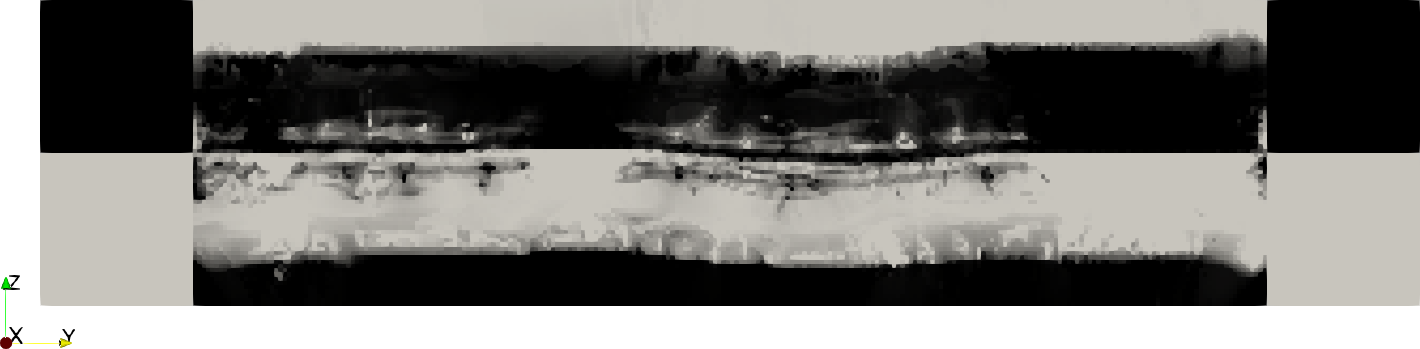}
    \caption{$\xi$}
    \label{fig:1varparchan_x}
    \end{subfigure}%
    ~
    \begin{subfigure}{\width}
    \includegraphics[width=\linewidth]{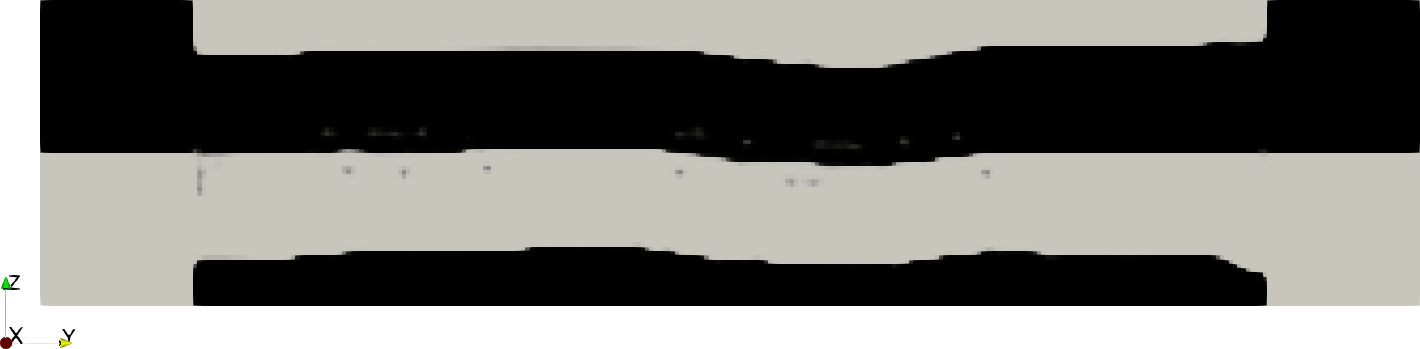}
    \caption{$\xip$}
    \label{fig:1varparchan_xp}
    \end{subfigure}
    \\
    \begin{subfigure}{\width}
    \includegraphics[width=\linewidth]{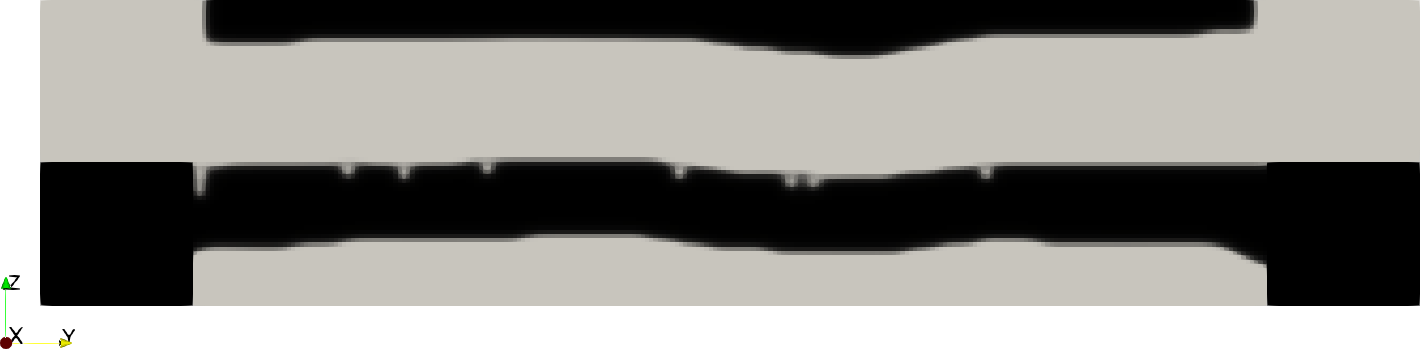}
    \caption{$\xiI$}
    \label{fig:1varparchan_x1}
    \end{subfigure}%
    ~
    \begin{subfigure}{\width}
    \includegraphics[width=\linewidth]{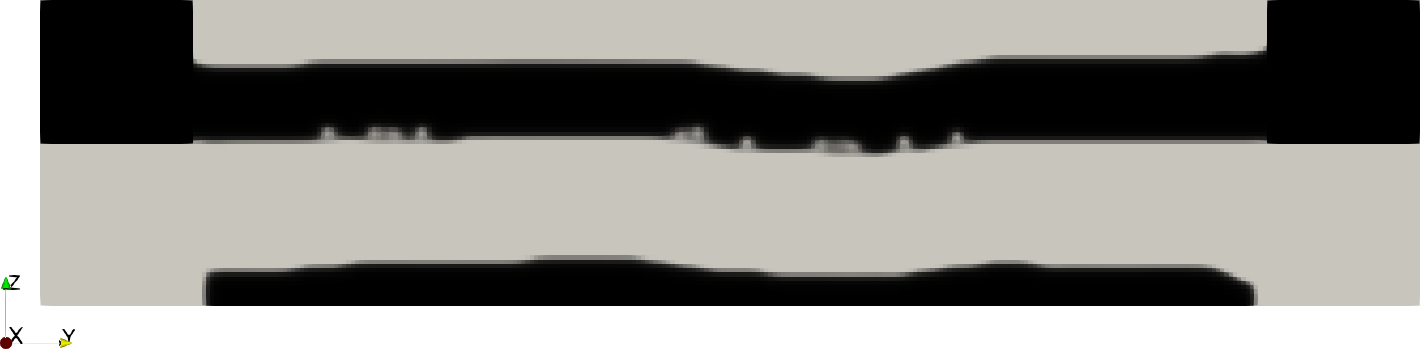}
    \caption{$\xiII$}
    \label{fig:1varparchan_x2}
    \end{subfigure}
    \\
    \begin{subfigure}{\width}
    \includegraphics[width=\linewidth]{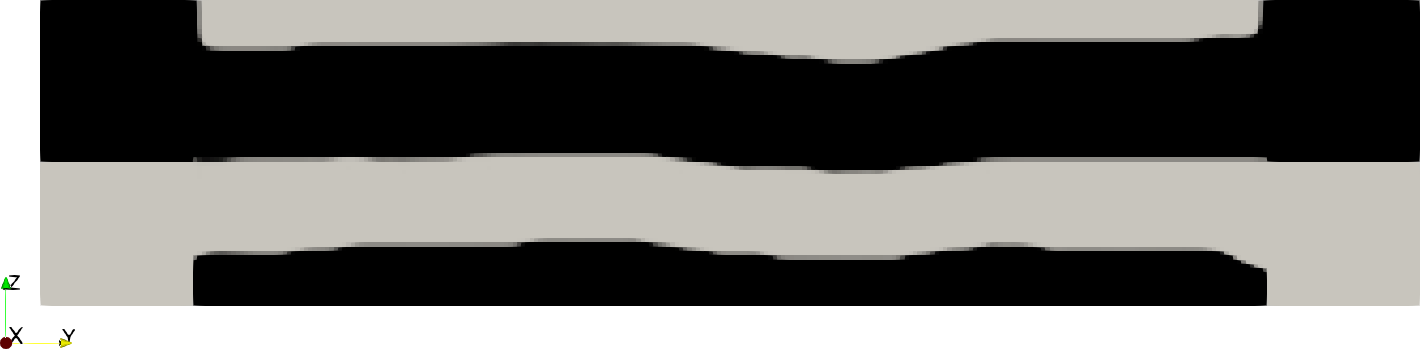}
    \caption{$\aI$}
    \label{fig:1varparchan_a1}
    \end{subfigure}%
    ~
    \begin{subfigure}{\width}
    \includegraphics[width=\linewidth]{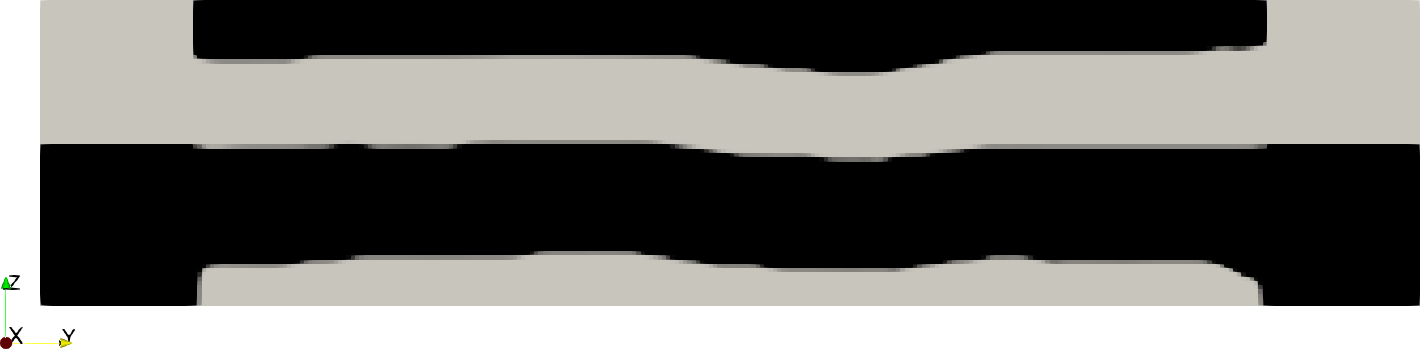}
    \caption{$\aII$}
    \label{fig:1varpar05_a2}
    \end{subfigure}
    \\
    \begin{subfigure}{\width}
    \includegraphics[width=\linewidth]{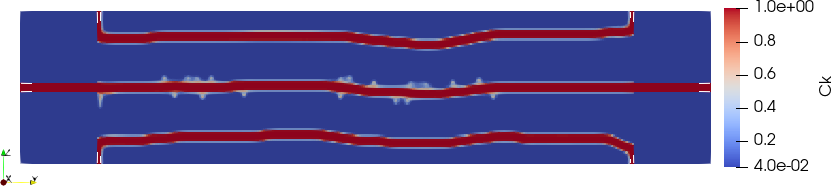}
    \caption{$\Ck$}
    \label{fig:1varparchan_Ck}
    \end{subfigure}
    \caption{Optimized fields, obtained by optimizing the parallel channels initial design, with a \dpf{5.0} pressure drop constraint. The design field $\xi$ (a) is filtered and projected to the filtered and projected field $\xih$ (b). The erosion-dilation process, discussed in Section \ref{sec:erodila}, is then used to generate $\xiI$ (c) and $\xiII$ (d), where the black region indicates the presence of the respective fluid. These fields lead to the physical fields (e-g), interpolated as discussed in Section \ref{sec:interpfunc}.}
    \label{fig:parchan_dp5fields}
\end{figure*}

\begin{figure}[tb]
    \centering
    \includegraphics[width=\linewidth]{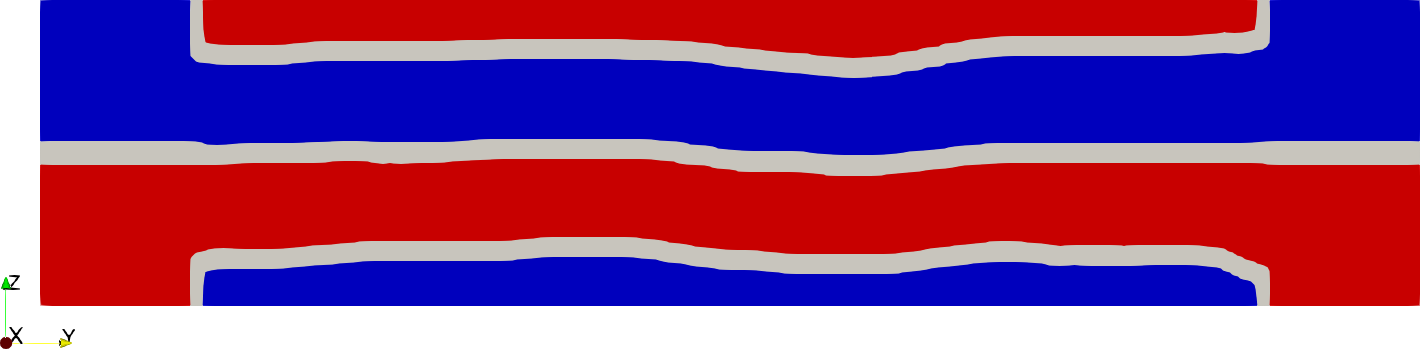}
    \caption{Overview of the fluid domains in the design optimized for \dpf{5.0}, with the parallel channels initial design. Red color denotes fluid 1, blue color fluid 2, and white the solid interface separating the two fluids.}
    \label{fig:2dfluiddomains}
\end{figure}

Figure \ref{fig:parchan_dp5fields} shows the design field $\xi$ and the intermediate fields, obtained by optimizing with an admissible pressure drop $\dpf{5}$ with the parallel channel initial design. From the eroded and dilated design fields, $\xiI$ and $\xiII$, seen in Figures \ref{fig:1varparchan_x1} and \ref{fig:1varparchan_x2}, it is seen that the fields have the same features, but that they are eroded and dilated versions of Figure \ref{fig:1varparchan_xp}, which results in the wall. It is observed that the islands at the top and bottom of the computational domain are of the opposite fluid. This is also seen in Figure \ref{fig:2dfluiddomains}, where the fluid domains are shown for the optimized heat exchanger design.
In Figure \ref{fig:1varparchan_Ck}, the relative conductivity, $\Ck$, in the design domain is shown. The walls are clearly identified with the higher conductivity. At the transition between the solid and the fluid, some few elements of intermediate conductivities are observed. However, this has very small influence on the final design, as very few elements are concerned. 

\begin{figure*}
    \centering
    \begin{subfigure}{\width}
    \includegraphics[width=\linewidth]{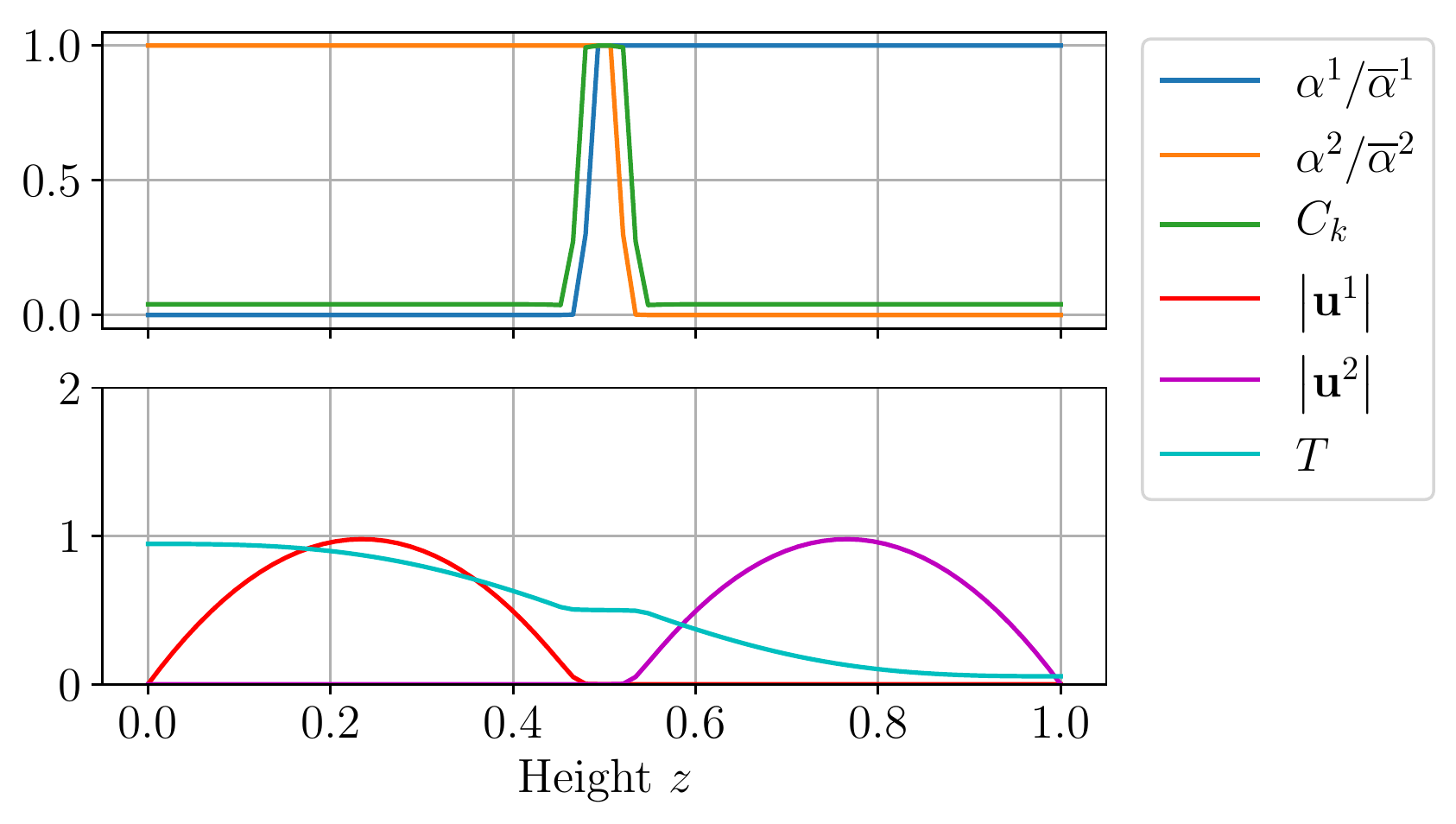}
    \caption{ }
    \label{fig:probebase}
    \end{subfigure}
    \begin{subfigure}{\width}
    \includegraphics[width=\linewidth]{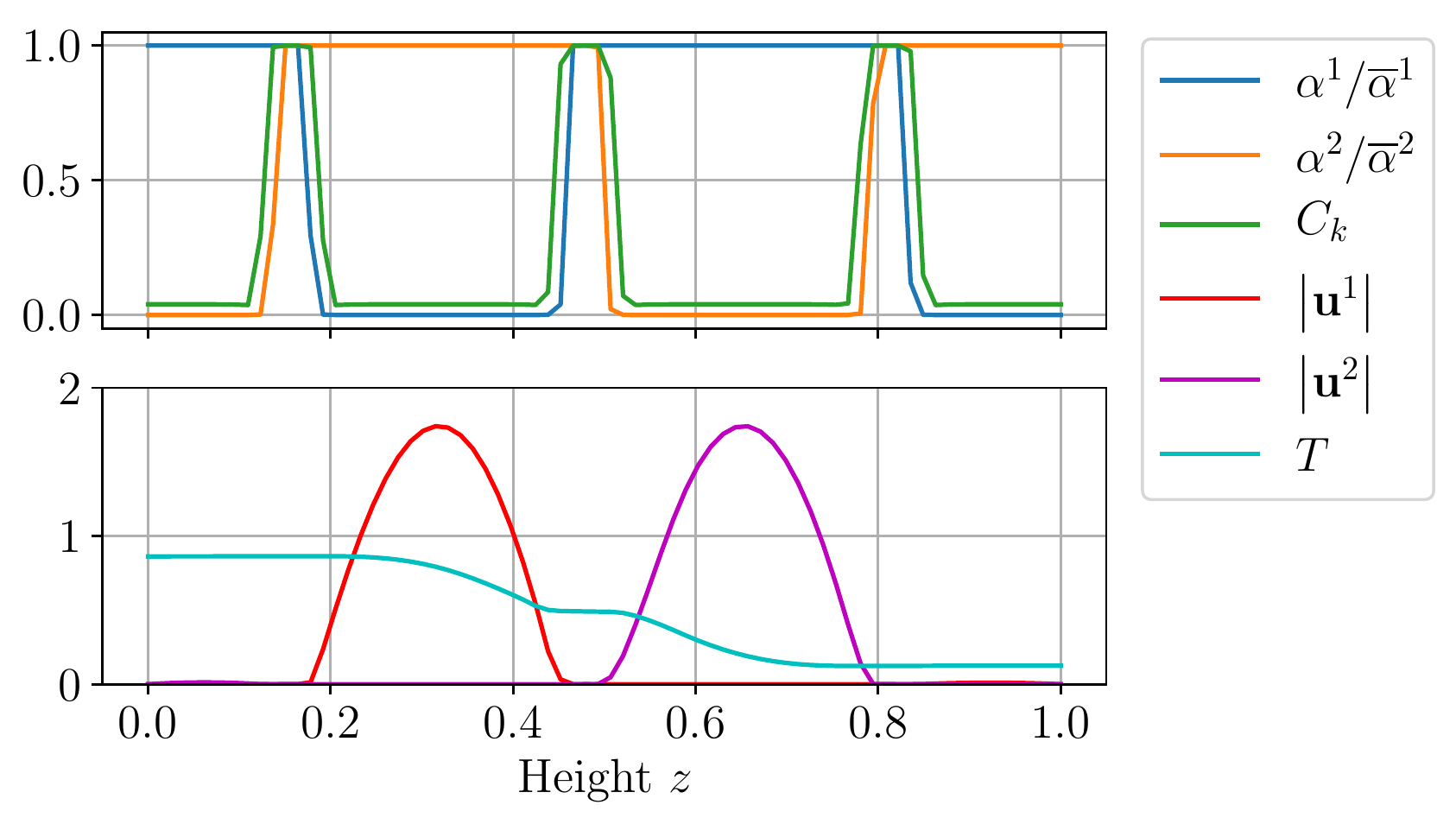}
    \caption{ }
    \label{fig:probe08}
    \end{subfigure}
    \caption{Probes showing physical quantities, velocity magnitudes and temperatures along the vertical centerline in the (a) baseline design and (b) design optimized with the initial channels and \dpf{5}.}
    \label{fig:parchanprobes}
\end{figure*}

In order to illustrate the distribution of the physical fields $\alpha^\gamma$ and $C_k$ along with their effect on the velocities and temperature, these quantities are probed along the vertical centerline in the baseline design seen in Figure \ref{fig:parchanbase} and in the design optimized with the baseline used as initial design and \dpf{5}, seen in Figure \ref{fig:1varparchan_8}. The probes are seen in Figure \ref{fig:parchanprobes}. In both cases, the wall is identified by the high relative conductivity and impermeabilities, as well as the low temperature gradient, caused by the high conductivities in the wall. In the probe of the optimized design, Figure \ref{fig:probe08}, the \textit{islands} of fluid are again highlighted. They seem to act as heat reservoirs. The flow of both fluids is seen to be present in the corresponding main fluid domain (no flow in the islands).

\begin{figure*}[htb]
    \centering
    \includegraphics[width=0.8\linewidth]{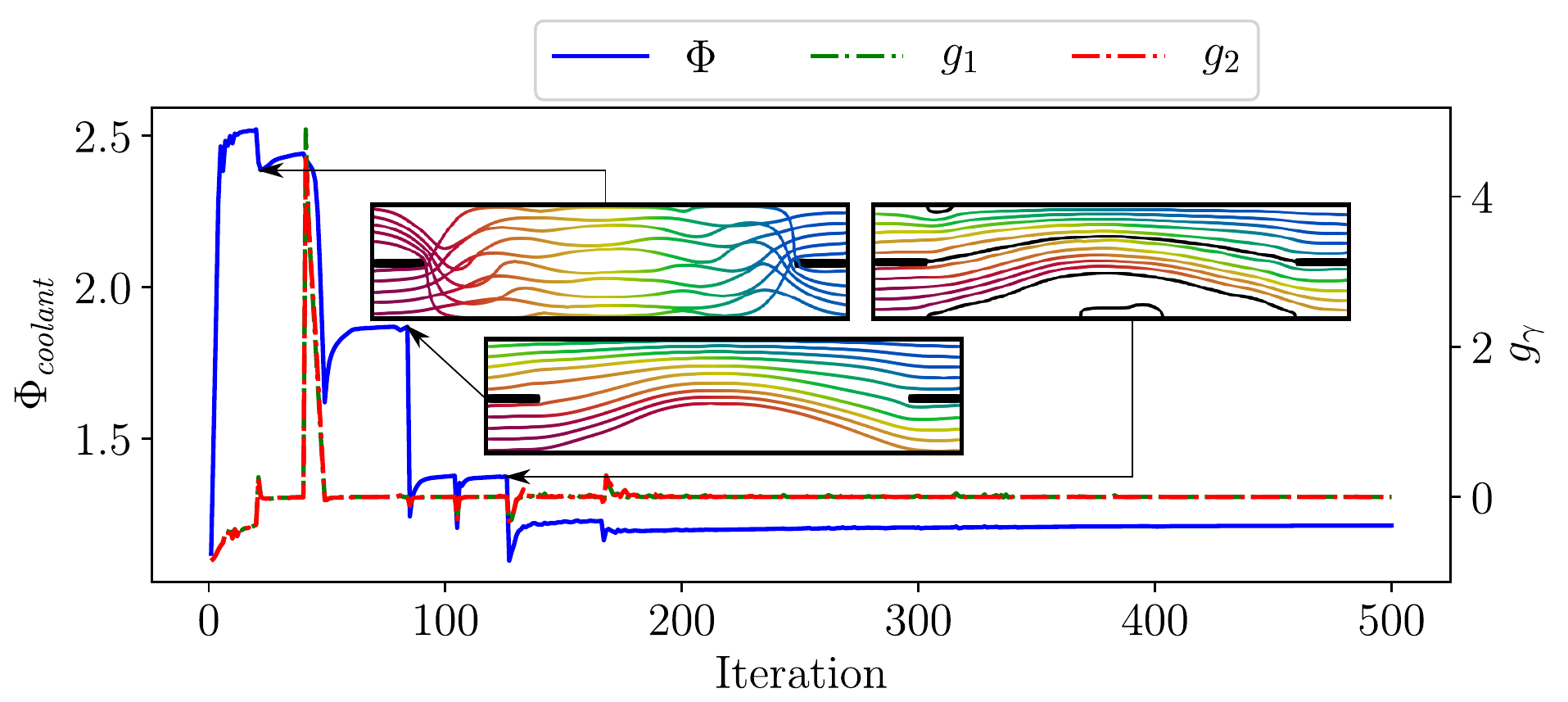}
    \caption{Objective- and constraint function, $\phih$ and $g_\gamma$, respectively, history over design iterations for the optimization with the initial channel design and \dpf{5}. Early design iterations, with low projection sharpness, show that no interface is formed and fluids are not well separated. As the continuation progresses, fluids get more separated, and finally, an interface is formed.}
    \label{fig:parchan_history}
\end{figure*}

The optimization history of the design with \dpf{5} and starting with $\xi=0.5$, is seen in Figure \ref{fig:parchan_history}. Here it is seen that the objective function is very low, near the theoretical maximum, in the early design iterations. The continuation steps are clearly identified, where spikes appear in the constraint function values and where the objective function value rises. The three shown preliminary designs from just prior to the continuation steps clearly illustrate that the two fluids are able to mix, when the projection value is low. This is due to the combination of the identification method discussed in Section \ref{sec:erodila} and the porous formulation. 
It is hence seen (from the first history state, where the streamlines mix) that the pressure constraint does not really have an influence before the continuation step to $\beta=4$, as the flow is mostly porous prior to this point. 

\subsection{Three-dimensional shell-and-tube heat exchanger}
\subsubsection{Problem definition}
\label{sec:probdef3d}
\begin{figure}[tbp]
    \centering
    \includegraphics[width=\linewidth]{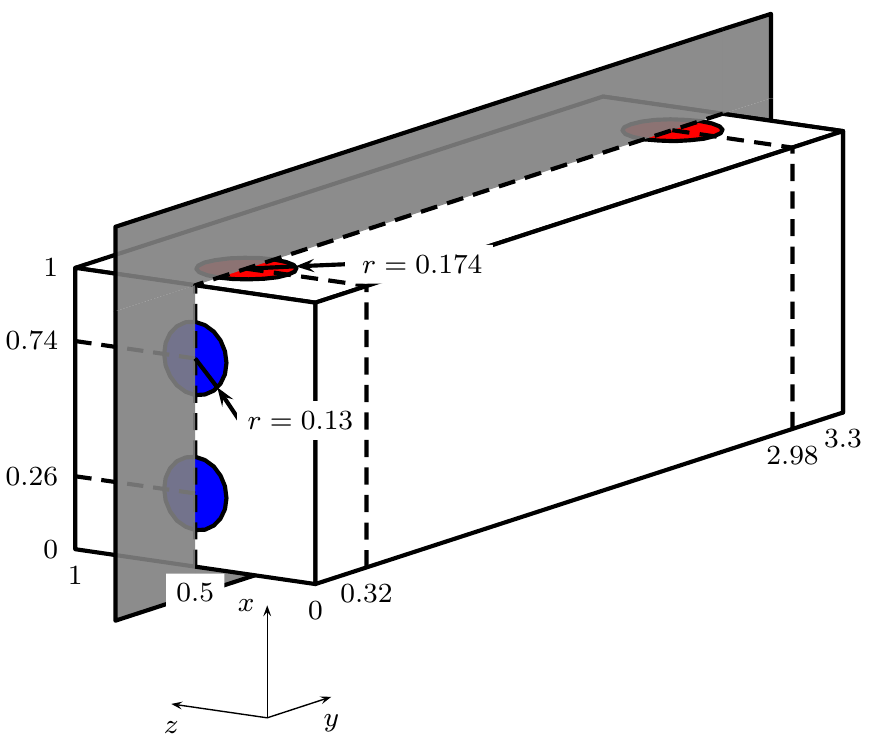}
    \caption{Design domain of the considered heat exchanger. The in- and outlets of the coolant, water, are located on the $y$-min plane (in blue) and the in- and outlets of the cooled fluid, oil, are located on the $x$-max plane (in red). The grey plane is the symmetry plane located at $z=0.5$. The reference length is the height of the heat exchanger, $L=146\cdot10^{-3}\;\mathrm{[m]}$.}
    \label{fig:3ddef}
\end{figure}

The second considered case is inspired by a shell-and-tube heat exchanger design. Figure \ref{fig:3ddef} shows the design problem. The heat exchanger has circular in- and outlets of the coolant on one face and circular in- and outlets of the cooled fluid on an adjacent face. Both inlet velocity profiles are assumed to be parabolic with respect to radial distance.

Based on the information, presented in Table \ref{tab:vestas_param}, the Reynolds- and Peclet numbers of both the cooled and coolant phase can be computed. The solid material is set to be stainless steel, the hot fluid to be oil and the coolant to be water.
\begin{table}[htbp]
    \centering
    \caption{Parameters of heat exchanger}
    \label{tab:vestas_param}
    \resizebox{\linewidth}{!}{\begin{tabular}{l||rr}
        \multirow{2}{*}{Property}                                                                & Cooled side      &  Coolant side\\
        & \multicolumn{1}{c}{Oil, 1} & \multicolumn{1}{c}{Water, 2}\\\hline\hline
        Mass flow $\scriptstyle\Dot{m}\;\mathrm{[kg\cdot s^{-1}]}$               & $4.57\cdot10^{-3}$        & $9.64\cdot10^{-4}$\\
        Inlet temperature $\scriptstyle T\;\mathrm{[C]}$                        & $90$          & $ 65$\\
        Density $\scriptstyle\rho\;\mathrm{[kg\cdot m^{-3}]}$                   & $866.4$       & $980$\\
        Dynamic viscosity $\scriptstyle\mu\;\mathrm{[N\cdot s\cdot m^{2}]}$     & $0.0215$      & $4.32\cdot10^{-4}$\\
        Heat capacity $\scriptstyle c_p\;\mathrm{[J\cdot kg^{-1}\cdot K^{-1}]}$ & $2088$        & $4188$\\
        Conductivity $\scriptstyle k\;\mathrm{[W\cdot m^{-1}\cdot K^{-1}]}$     & $0.1233$      & $0.6$\\
        Solid conductivity $\scriptstyle k\;\mathrm{[W\cdot m^{-1}\cdot K^{-1}]}$     & \multicolumn{2}{c}{$30$}\\
        Pressure drop $\scriptstyle\Delta p\;\mathrm{[Pa]}$                     & $16402$       & $3080$\\
        Reynolds number $\scriptstyle[-]$                                       & $10.9$        & $150$\\
        Peclet number $\scriptstyle[-]$                                         & $3973$        & $453$
    \end{tabular}}
\end{table}
The Reynolds and solid Peclet numbers seen in Table \ref{tab:vestas_param} are both converted to their computational equivalents, adjusting for the non-dimensional inlet diameters \cite{Alexandersen2016}. The Darcy number is set to $\mathit{Da}=10^8$ to limit the porous flow. The final computational parameters are shown in Table \ref{tab:3D_param}.
\begin{table}[tbp]
    \centering
    \caption{Computational parameters of the three dimensional heat exchanger optimization problem.}
    \label{tab:3D_param}
    \begin{tabular}{l||rr}
        Property                                & Cooled      &  Coolant\\\hline\hline
        Reynolds number $\scriptstyle\Re[-]$       &    $31.32$     &  $578$\\
        Solid Peclet number $\scriptstyle\Pes[-]$         & $46.92$        & $34.8$\\
        Conductivity ratio $\scriptstyle\Ck[-]$ & $4.11\cdot10^{-3}$        & $0.02$\\
        Impermeability $\scriptstyle\amax[-]$ & $3.19\cdot10^6$ & $1.73\cdot10^5$
    \end{tabular}
\end{table}

\begin{figure*}[hbt]
    \centering
    \begin{subfigure}{\width}
    \includegraphics[width=\linewidth]{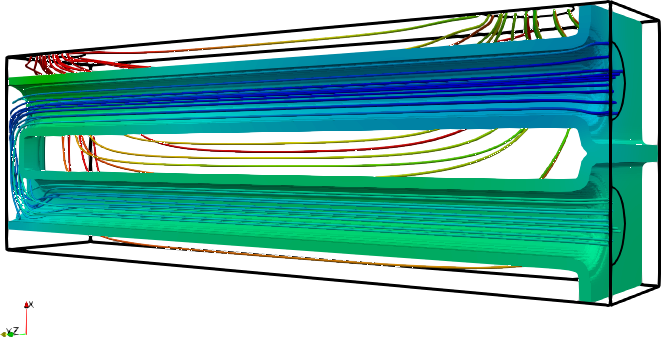}
    \caption{One tube}
    \label{fig:3Dbase1}
    \end{subfigure}
    ~
    \begin{subfigure}{\width}
    \includegraphics[width=\linewidth]{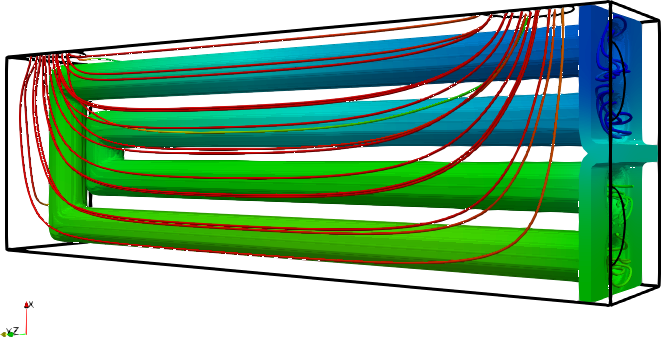}
    \caption{Four tubes}
    \label{fig:3Dbase2}
    \end{subfigure}
    \caption{Half baseline designs, with the symmetry plane depicted in Figure \ref{fig:3ddef}. Structures and streamlines are colored by temperature, in- and outlets highlighted with black contours, and structure thresholded at $\aI>0.1\amI\land\aII>0.1\amII$.}
    \label{fig:3Dbase}
\end{figure*}

The optimization is carried out on the half domain, as depicted by the $z=0.5$ plane in Figure \ref{fig:3ddef}. The boundary conditions on this plane are of symmetric type, with no-through flux of mass and heat. This symmetric boundary condition constrains the obtained designs to be symmetric. Two baseline designs are considered, both inspired by shell-and-tube heat exchangers. In both cases, the coolant fluid is the tube side. The coolant fluid flows through one or four tubes (respectively half and two tubes in the half domain), in the two baseline designs seen in Figure \ref{fig:3Dbase}, respectively. Near the plane with the coolant inlet and outlet, a manifold is located which redirects the coolant fluid and a small wall separates the coolant in- and outlet manifolds.

The baseline designs have different pressure drops and heat transfers, seen in Table \ref{tab:3dbase}.
\begin{table}[tbp]
    \centering
    \caption{Comparison of performance of the two baseline designs.}
    \label{tab:3dbase}
    \begin{tabular}{l||rrr}
        Design & $\Delta P^1_\mathit{base}$ & $\Delta P^2_\mathit{base}$ & $\Phi_\mathit{coolant,\,base}$\\\hline\hline
        1 tube & $11.9$ & $3.56$ & $4.68$\\
        4 tubes & $57.4$ & $3.08$ & $7.27$ 
    \end{tabular}
\end{table}
The baseline designs are optimized with their corresponding pressure drops, i.e. $\dpmax^\gamma=\dpbase^\gamma$. For each baseline pressure drop, four initial designs are considered: One where the design variable is uniform $\xi=0.5$ in the design domain, and three where the respective baseline design is utilized, but subjected to different projection sharpness, $\beta_\mathit{initial}=\{1,\;4,\;8\}$. In all cases, the optimization is cut-off after 350 design iterations.
The higher initial $\beta$ values ensures more well-defined and impermeable walls. This enables the the coolant flow to travel through the initial tubes, reaching deeper into the domain in the early design iterations. However, for lower $\beta$ values and the uniform distribution of $\xi=0.5$, the walls are either fairly porous or non-existent. This means that the coolant will take the path of least resistance, which is directly from inlet to outlet and not reaching deep into the computational domain. Having the flow fields reaching far into the design domain yields a wider range of spatial sensitivity information. Hence, the flow field in the initial design, and the early design iterations, has a very high impact on the final design and will vary depending on the given problem and boundary conditions.

\subsubsection{Results}
\label{sec:res1var3d}
\begin{figure*}[htbp]
    \centering
    \begin{subfigure}{\width}
    \includegraphics[width=\linewidth]{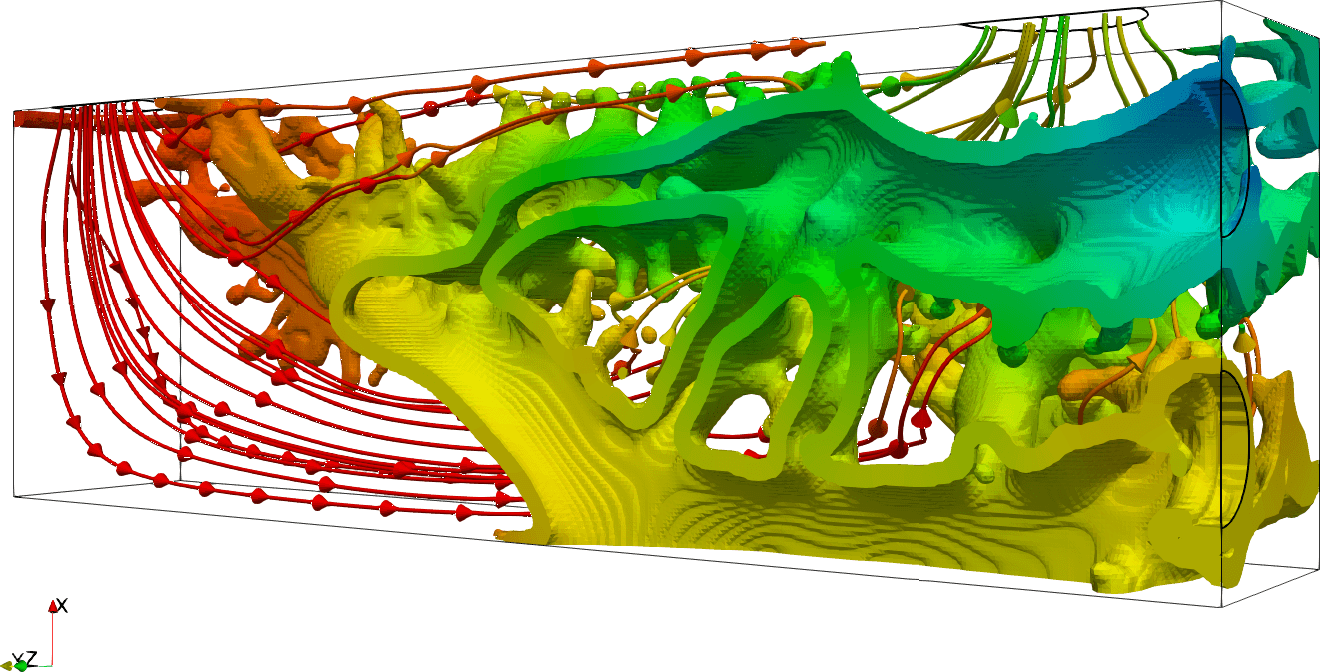}
    \caption{One channel baseline, starting with $\xi=0.5$ and $\bini=1$, \obj{12.71}}
    \label{fig:structure_base1_res}
    \end{subfigure}
    ~
    \begin{subfigure}{\width}
    \includegraphics[width=\linewidth]{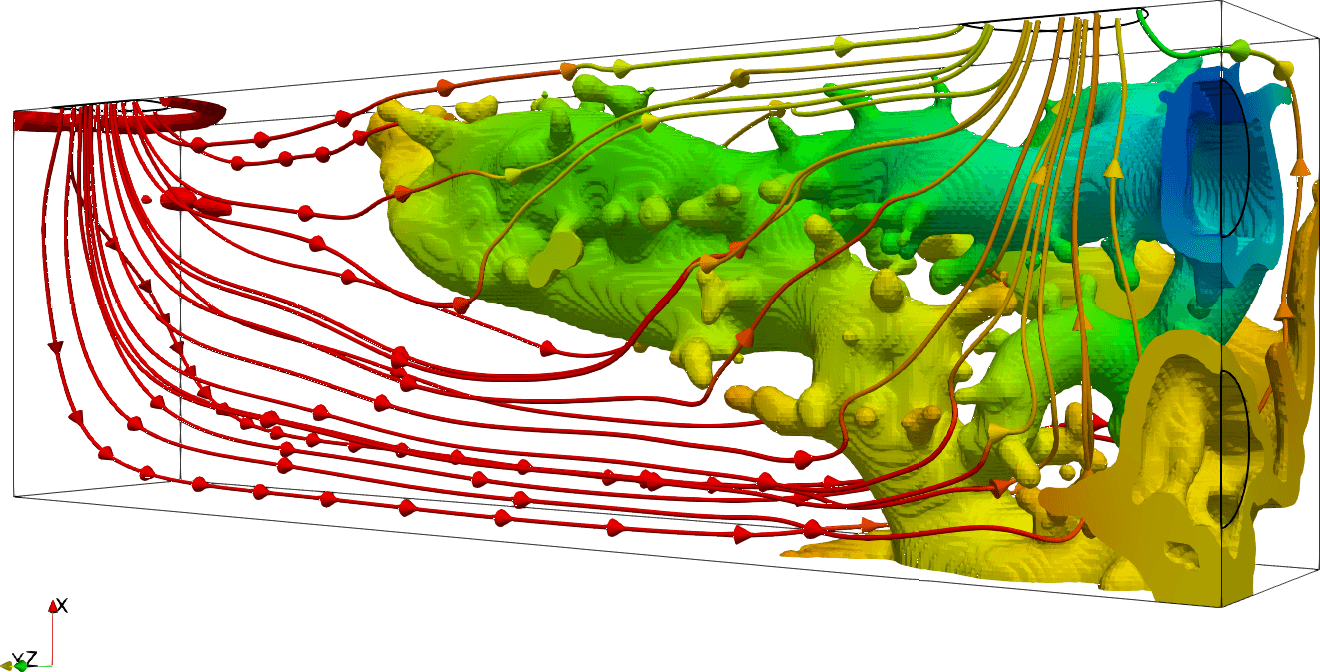}
    \caption{Four channel baseline, starting with $\xi=0.5$ and $\bini=1$, \obj{13.69}}
    \label{fig:structure_base2_res}
    \end{subfigure}
    \\
    \begin{subfigure}{\width}
    \includegraphics[width=\linewidth]{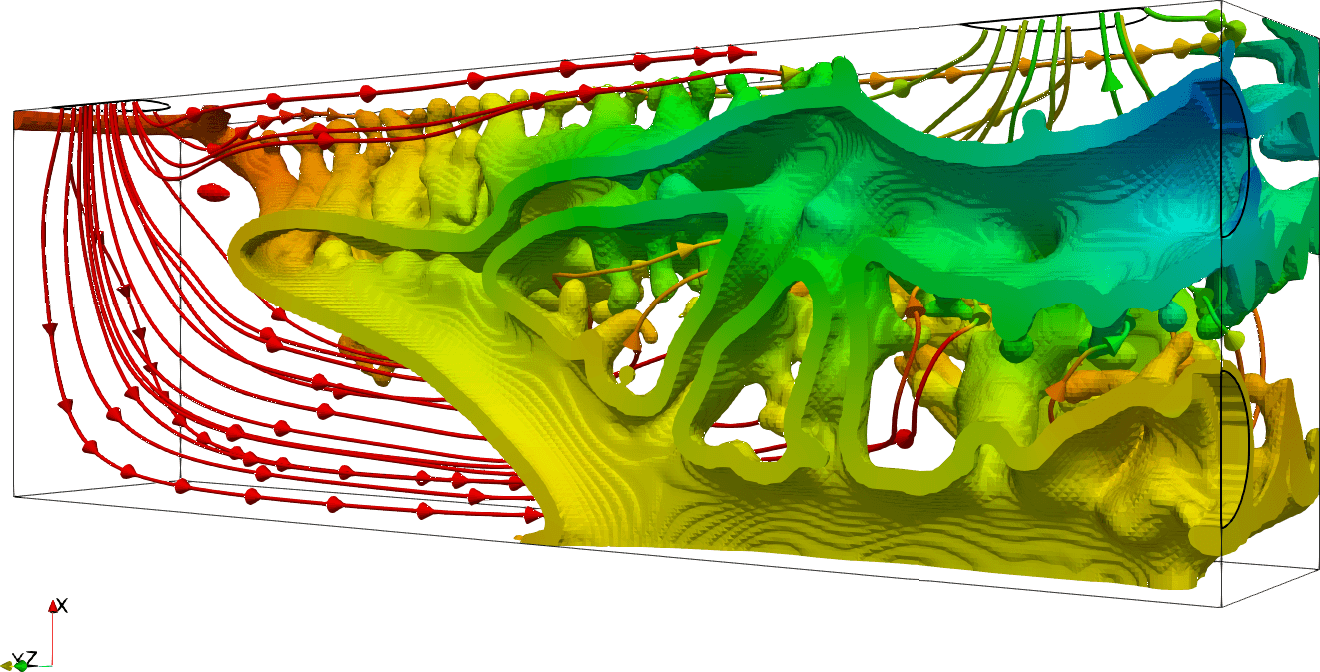}
    \caption{One channel baseline, starting from baseline with $\bini=1$, \obj{13.31}}
    \label{fig:structure_betas1_base1}
    \end{subfigure}
    ~
    \begin{subfigure}{\width}
    \includegraphics[width=\linewidth]{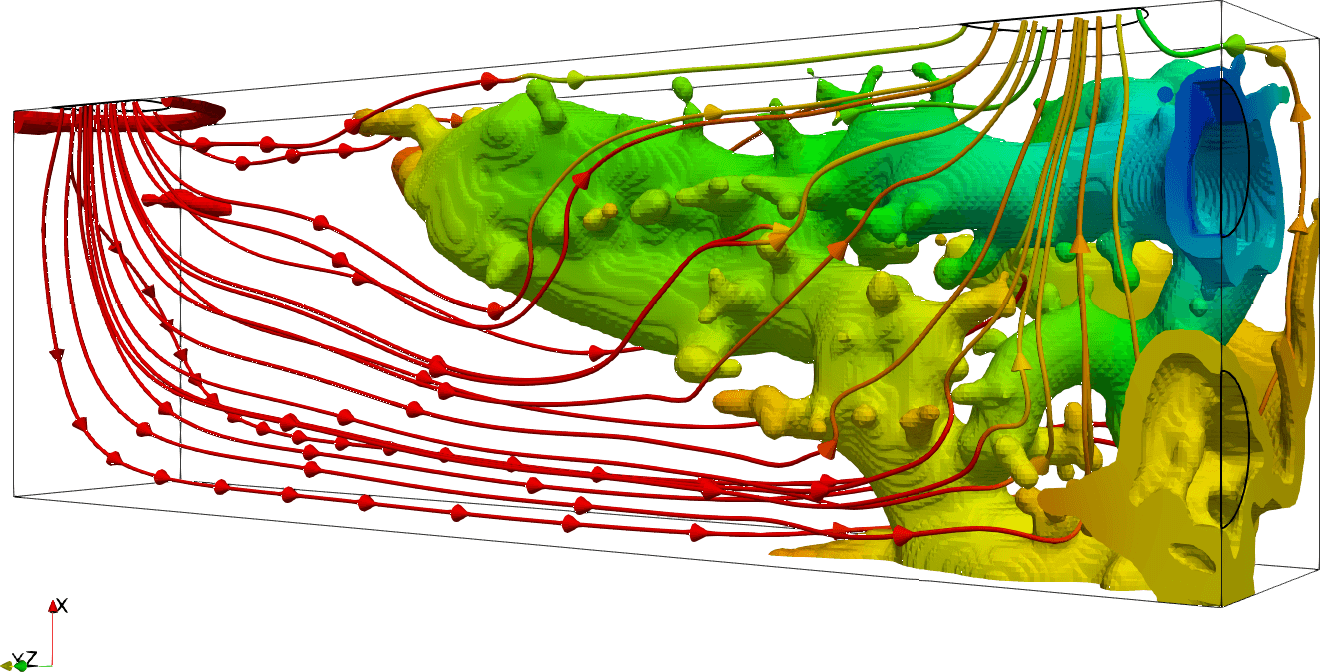}
    \caption{Four channel baseline, starting from baseline with $\bini=1$, \obj{13.80}}
    \label{fig:structure_betas1_base2}
    \end{subfigure}
    \\
    \begin{subfigure}{\width}
    \includegraphics[width=\linewidth]{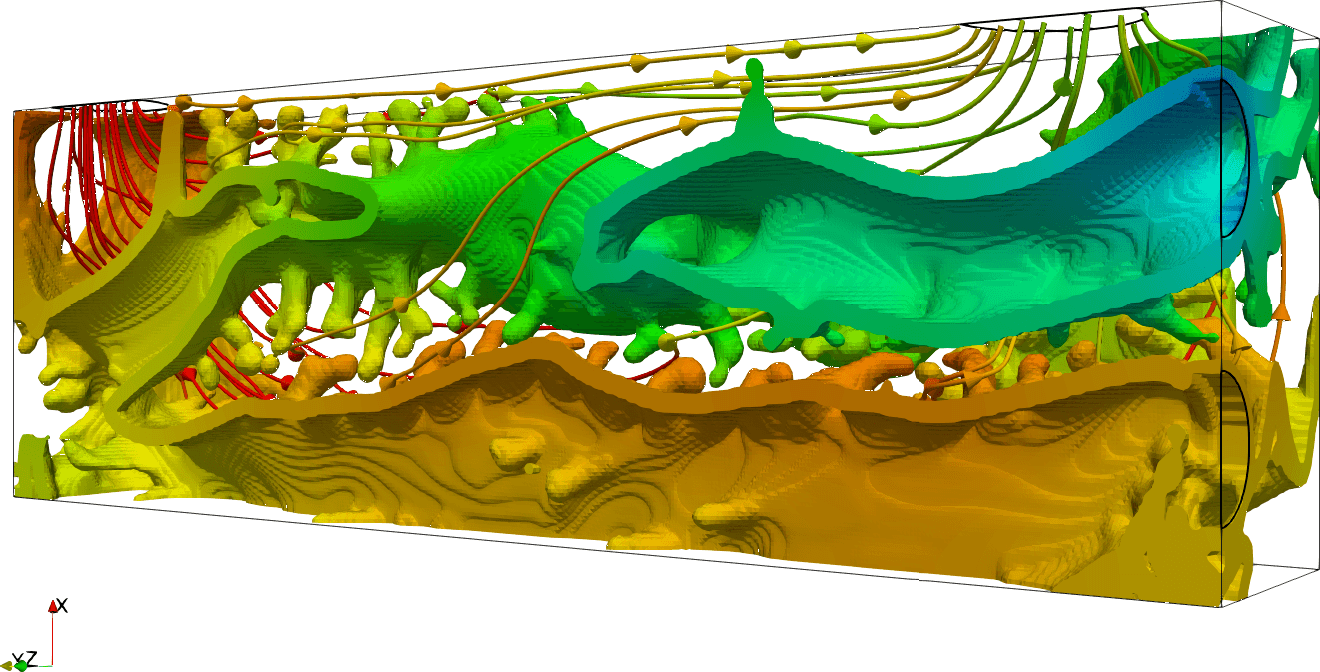}
    \caption{One channel baseline, starting from baseline with $\bini=4$, \obj{13.44}}
    \label{fig:structure_betas4_base1}
    \end{subfigure}
    ~
    \begin{subfigure}{\width}
    \fbox{
    \includegraphics[width=\linewidth]{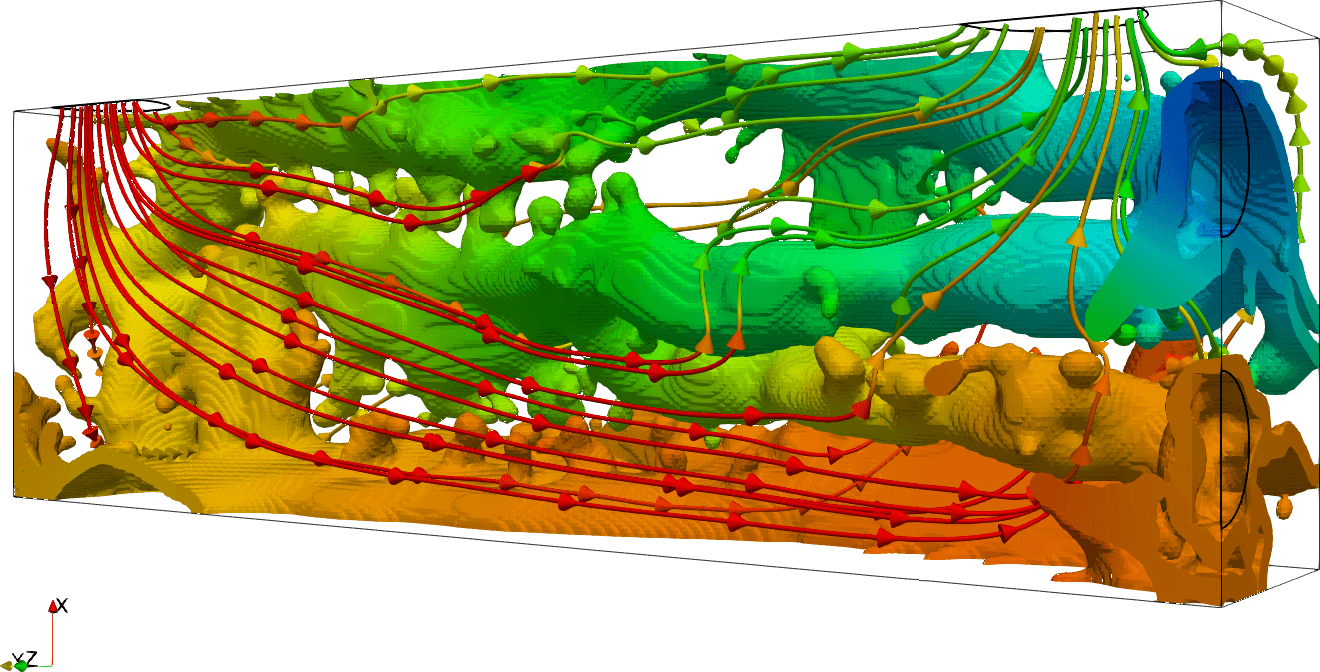}}
    \caption{Four channel baseline, starting from baseline with $\bini=4$, \obj{15.49}}
    \label{fig:structure_betas4_base2}
    \end{subfigure}
    \\
    \begin{subfigure}{\width}
    \includegraphics[width=\linewidth]{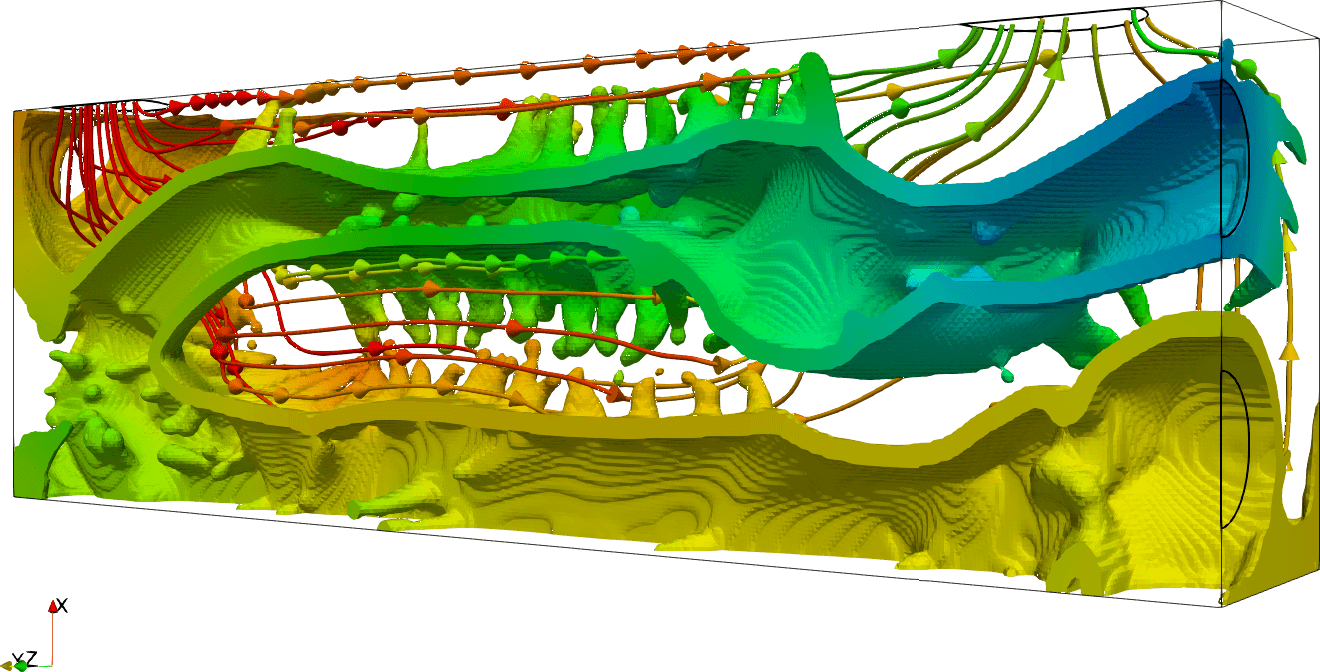}
    \caption{One channel baseline, starting from baseline with $\bini=8$, \obj{13.34}}
    \label{fig:structure_betas8_base1}
    \end{subfigure}
    ~
    \begin{subfigure}{\width}
    \includegraphics[width=\linewidth]{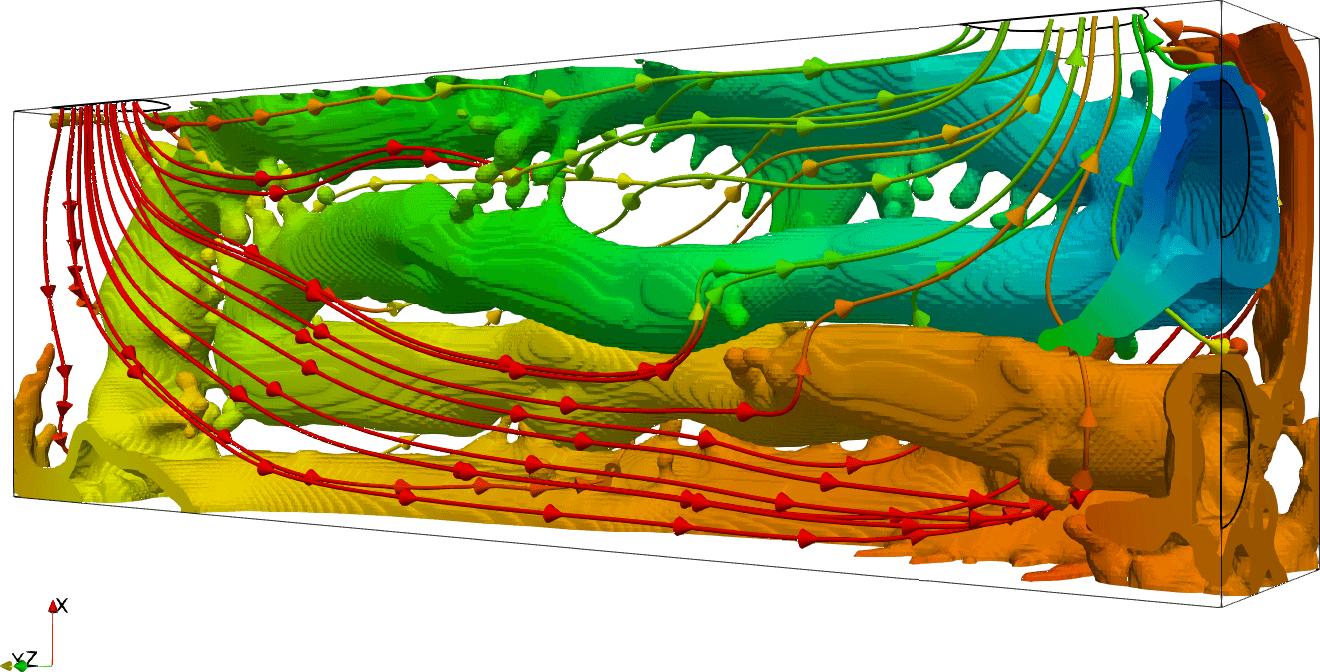}
    \caption{Four channel baseline, starting from baseline with $\bini=8$, \obj{15.16}}
    \label{fig:structure_betas8_base2}
    \end{subfigure}
    \\
    \begin{subfigure}{0.3\textwidth}
    \includegraphics[width=\linewidth]{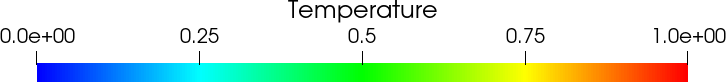}
    \end{subfigure}
    \caption{Optimized structures, and corresponding streamlines of the cooled fluid, both colored by temperature. The structures are thresholded at $\aI>0.1\amI\land\aII>0.1\amII$.}
    \label{fig:3dstructures}
\end{figure*}

With both baseline designs, seen in Figure \ref{fig:3Dbase}, the heat exchanger is optimized using the respective initial baseline designs and $\beta$ values, as previously discussed. All designs are optimized with the maximum pressure drop being equal to the pressure drop in its respective baseline design. In total, eight optimized designs are obtained and the solid structures with streamlines of the designs are seen in Figure \ref{fig:3dstructures}.
For the designs optimized from a loosely-defined initial design ($\bini=1$), seen in Figures \ref{fig:structure_base1_res}-\ref{fig:structure_betas1_base2}, the coolant channels do not reach the end of the design domain. However, as seen in Figures \ref{fig:structure_betas4_base1}-\ref{fig:structure_betas8_base2}, when starting from more well-defined initial designs (higher $\bini$), the coolant channels reach further into the design domain, 

\begin{figure}[htbp]
    \centering
    \includegraphics[width=\linewidth]{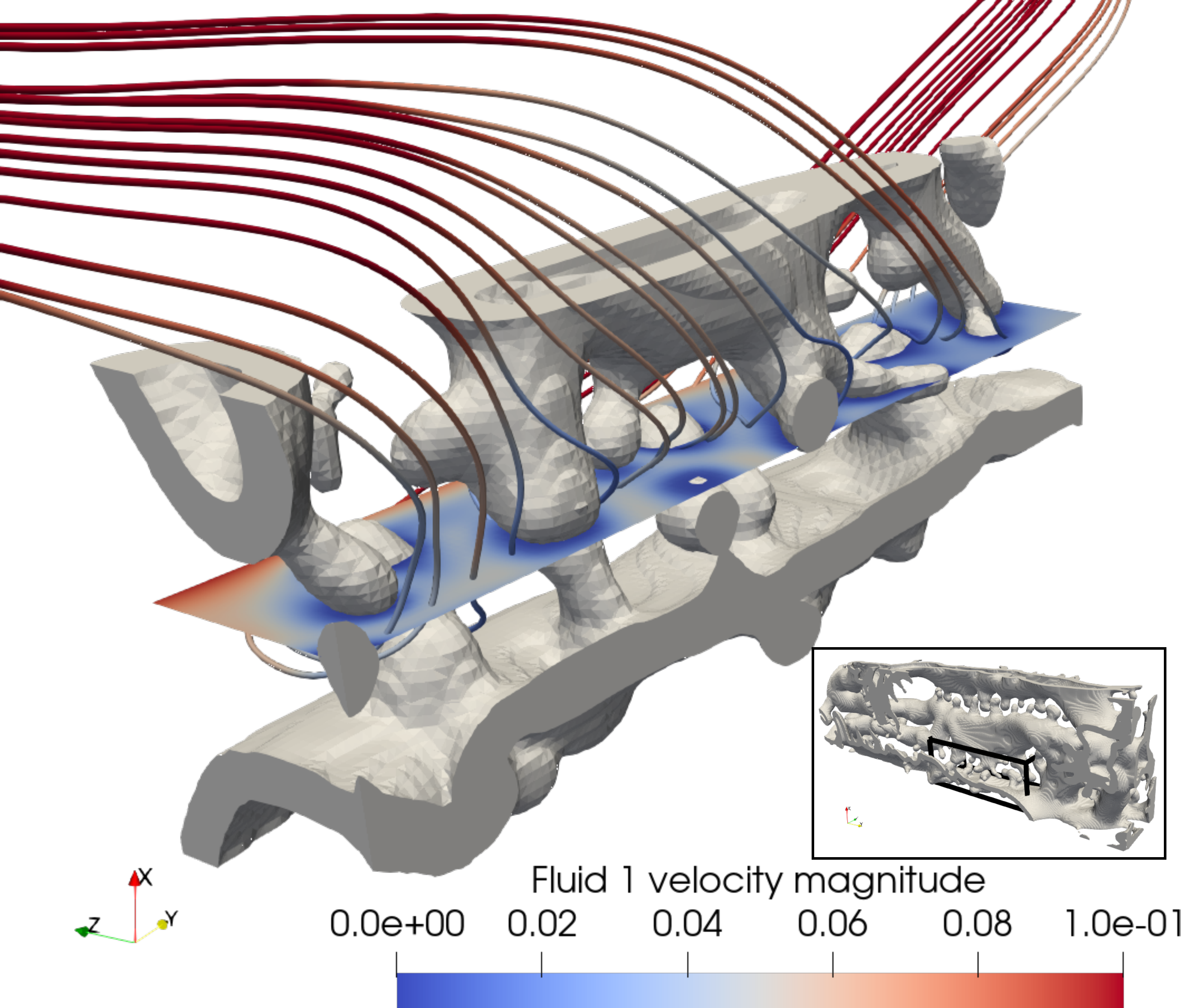}
    \caption{Cut of the solid structure of the design optimized with the four channel baseline, with the initial projection sharpness $\beta_\mathit{initial}=4$.The velocity magnitude of the cooled fluid 2 is seen on the plane. }
    \label{fig:extended_features}
\end{figure}

On the obtained solid structures, microvilli-like extended features are observed, going into the cooled fluid domain seemingly normal from the coolant tube surface. Microvilli are biological features, which are a part of the cellular membrane, increasing the surface area for absorption and other processes, but holding the volume increase to a minimum. The microvilli-like features from the different tubes are shifted in positions, as seen in the closeup from Figure \ref{fig:extended_features} (from the design seen in Figure \ref{fig:structure_betas4_base2}, optimized with $\beta_\mathit{initial}=4$ and the four channel baseline initial design field). It is seen that the microvilli-like features are placed in shifted positions relative to each other, such that the flow passes them in a curved manner, which increases the surface area, with flow next to it.

\begin{figure*}[htbp]
    \centering
    \begin{subfigure}{\width}
    \includegraphics[width=\linewidth]{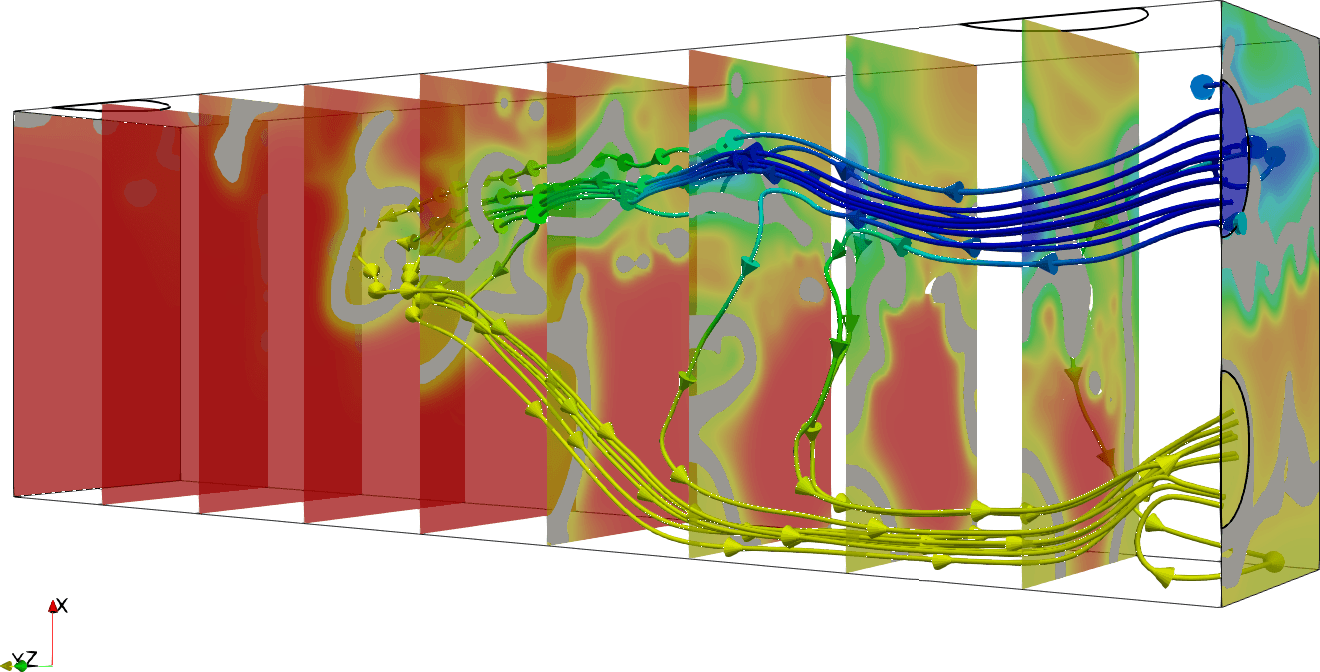}
    \caption{One channel baseline, starting with $\xi=0.5$ and $\bini=1$, \obj{12.71}}
    \label{fig:slices_base1_res}
    \end{subfigure}
    ~
    \begin{subfigure}{\width}
    \includegraphics[width=\linewidth]{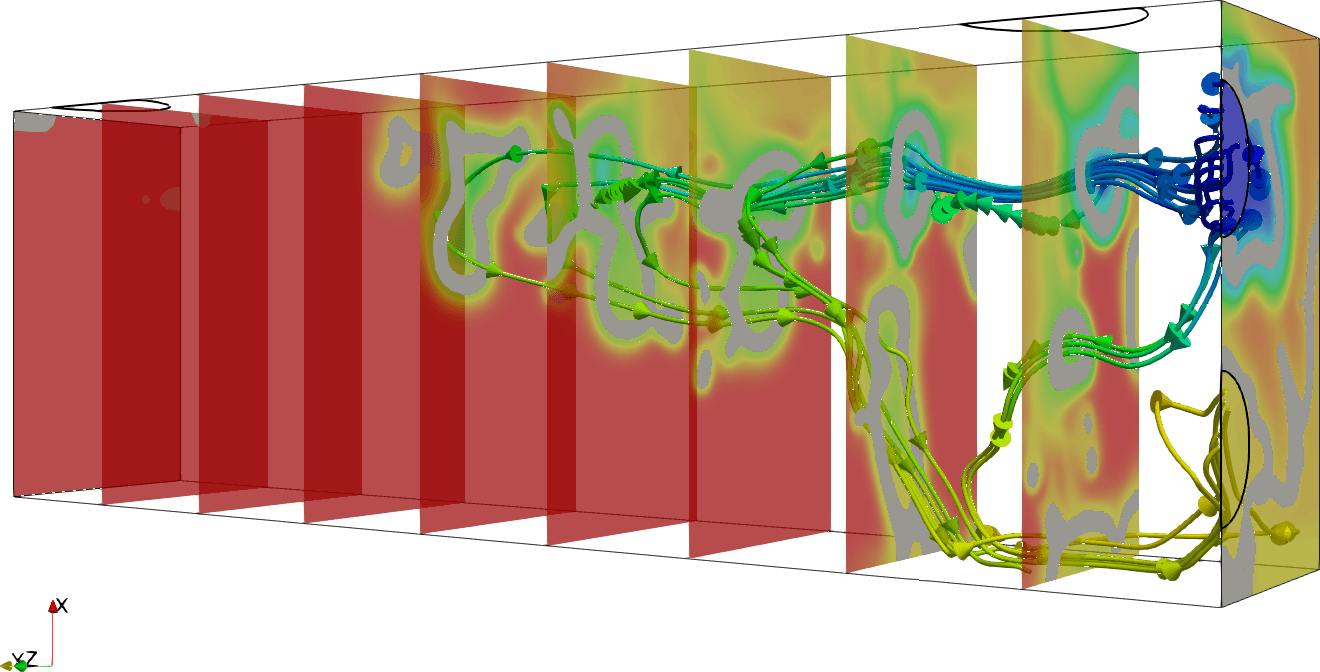}
    \caption{Four channel baseline, starting with $\xi=0.5$ and $\bini=1$, \obj{13.69}}
    \label{fig:slices_base2_res}
    \end{subfigure}
    \\
    \begin{subfigure}{\width}
    \includegraphics[width=\linewidth]{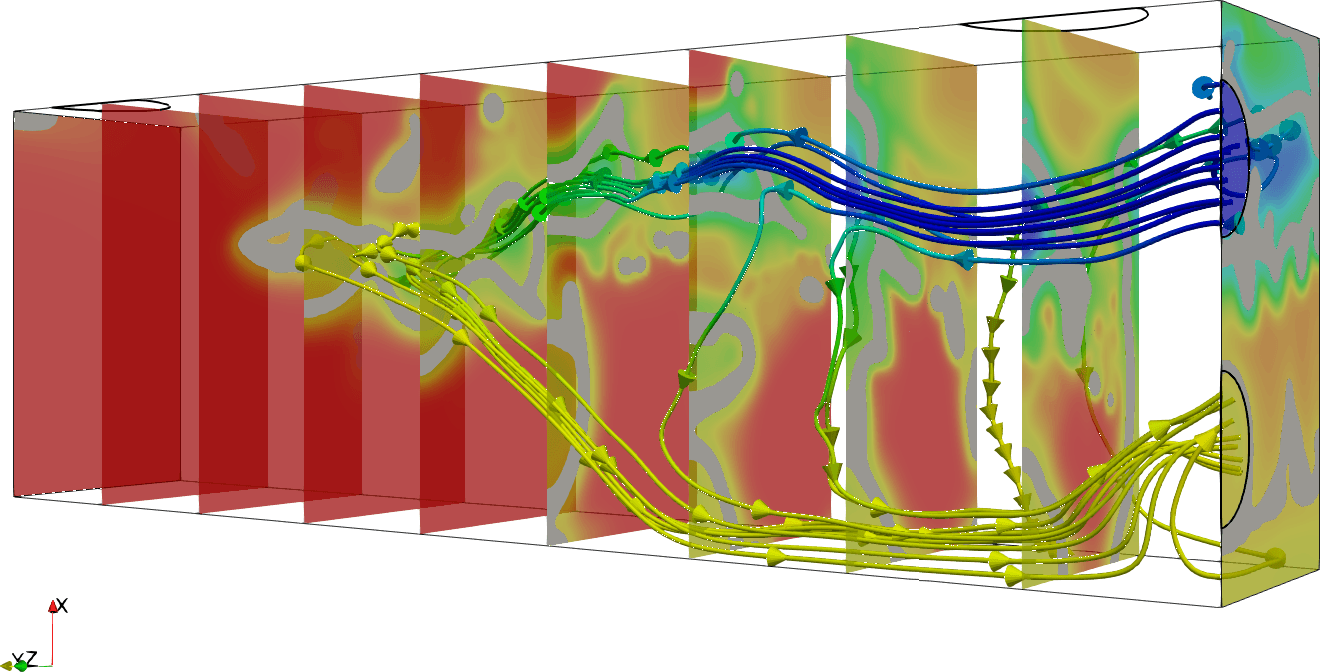}
    \caption{One channel baseline, starting from baseline with $\bini=1$, \obj{13.31}}
    \label{fig:slices_betas1_base1}
    \end{subfigure}
    ~
    \begin{subfigure}{\width}
    \includegraphics[width=\linewidth]{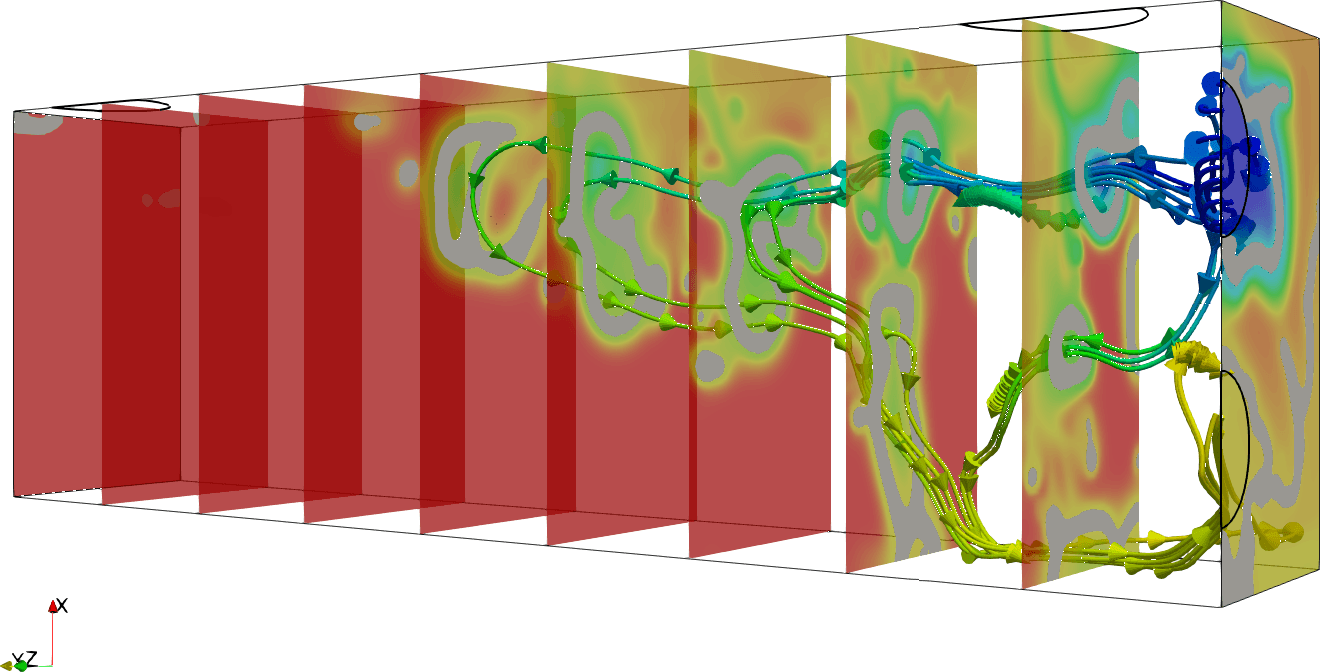}
    \caption{Four channel baseline, starting from baseline with $\bini=1$, \obj{13.80}}
    \label{fig:slices_betas1_base2}
    \end{subfigure}
    \\
    \begin{subfigure}{\width}
    \includegraphics[width=\linewidth]{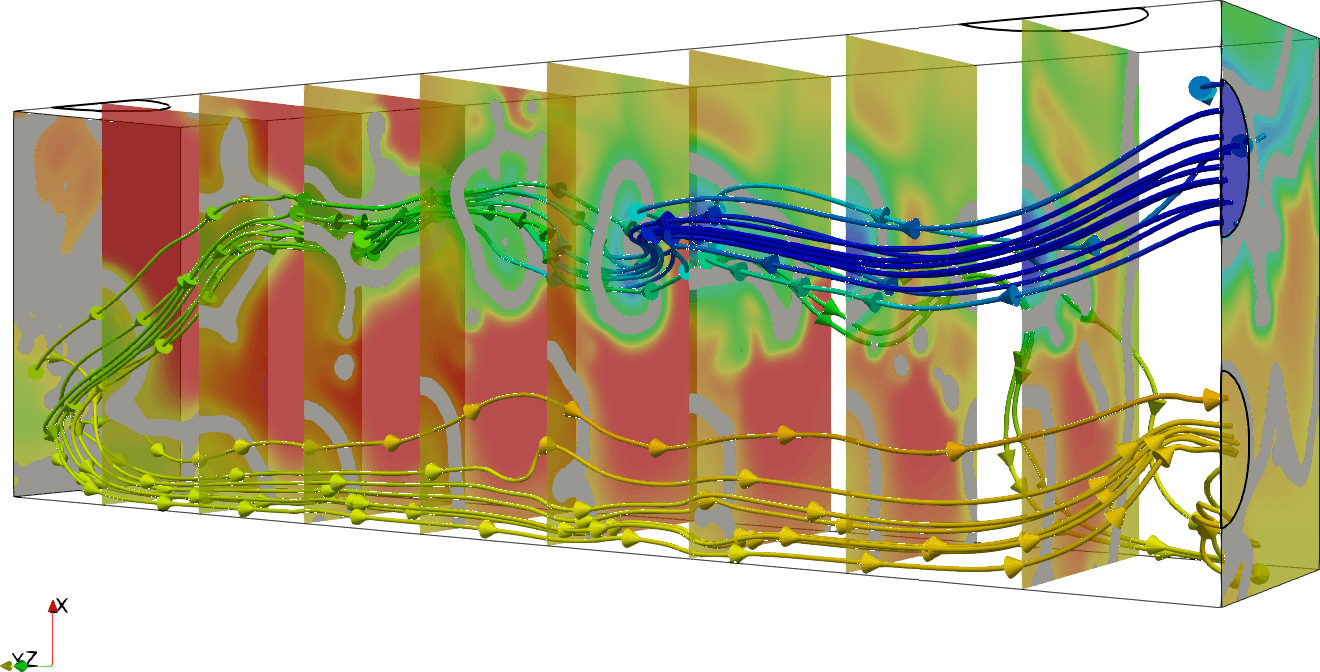}
    \caption{One channel baseline, starting from baseline with $\bini=4$\obj{13.44}}
    \label{fig:slices_betas4_base1}
    \end{subfigure}
    ~
    \begin{subfigure}{\width}
    \fbox{
    \includegraphics[width=\linewidth]{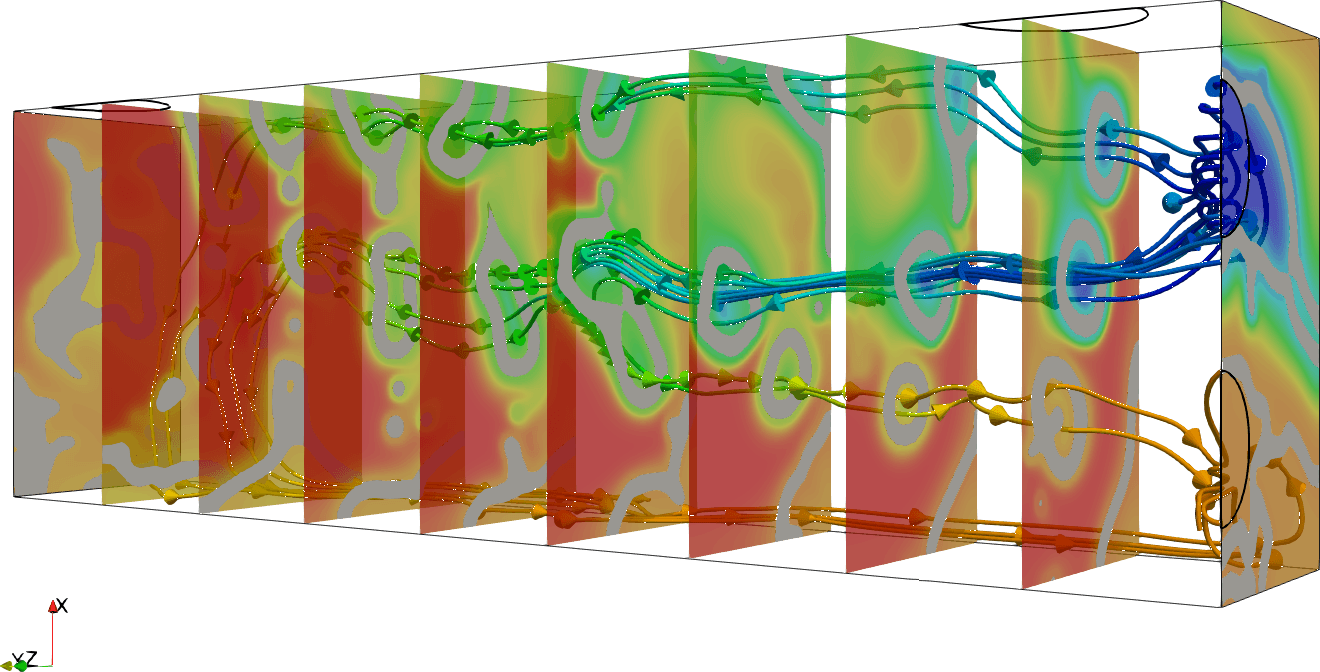}}
    \caption{Four channel baseline, starting from baseline with $\bini=4$, \obj{15.49}}
    \label{fig:slices_betas4_base2}
    \end{subfigure}
    \\
    \begin{subfigure}{\width}
    \includegraphics[width=\linewidth]{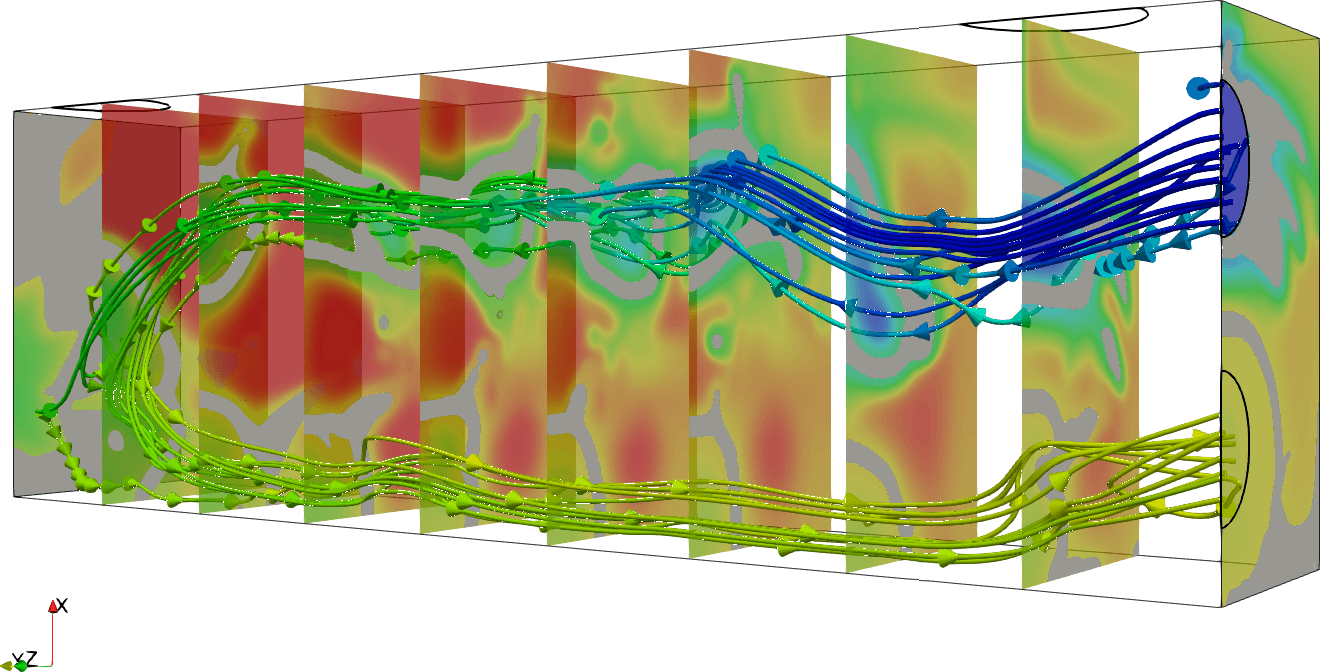}
    \caption{One channel baseline, starting from baseline with $\bini=8$, \obj{13.34}}
    \label{fig:slices_betas8_base1}
    \end{subfigure}
    ~
    \begin{subfigure}{\width}
    \includegraphics[width=\linewidth]{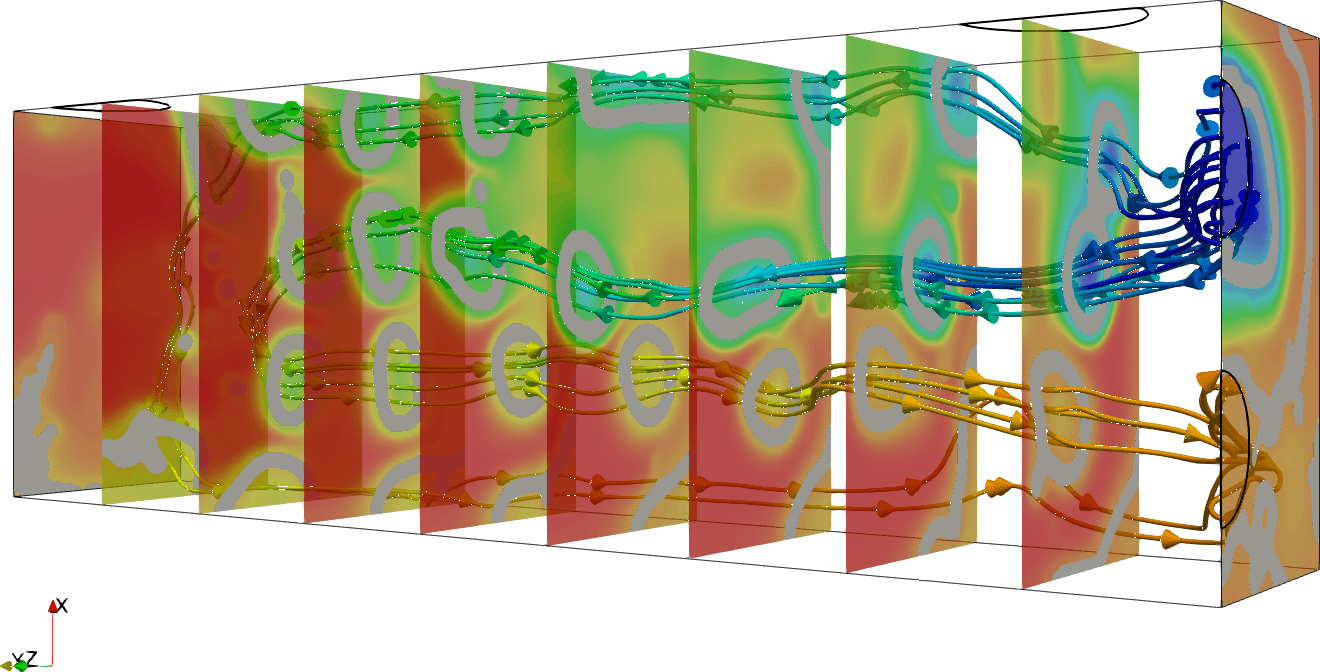}
    \caption{Four channel baseline, starting from baseline with $\bini=8$, \obj{15.16}}
    \label{fig:slices_betas8_base2}
    \end{subfigure}
    \\
    \begin{subfigure}{0.3\textwidth}
    \includegraphics[width=\linewidth]{vestas_temperature_legend.png}
    \end{subfigure}
    \caption{Streamlines of the coolant fluid and slices through the design domain. Slices and streamlines are colored by temperature. The structures (thresholded at $\aI>0.1\amI\land\aII>0.1\amII$) on the slices are colored in grey.}
    \label{fig:3dslices}
\end{figure*}
Figure \ref{fig:3dslices} shows the flow of the tube side coolant. It is again observed that starting with an initial design, where the coolant flow reaches the back of the domain, results in designs, where the coolant channels are present in the entirety of the domain. However, for the designs where the coolant flow did not reach far into the computational domain, Figures \ref{fig:slices_base1_res}-\ref{fig:slices_betas1_base2}, it is seen that the back of the domain is almost unused, as neither structures nor visible temperature variations are present in these areas. The entirety of the design domain is hence not fully exploited when starting with a too poorly-defined and permeable initial design.
On the other hand, when a higher initial projection sharpness is used, the topology of the initial design of the coolant fluid is preserved to a certain extent, as shown in Figures \ref{fig:slices_betas4_base1}-\ref{fig:slices_betas8_base2}. The designs obtained with $\bini>1$ and the one channel baseline design, seen in Figures \ref{fig:structure_betas4_base1}, \ref{fig:structure_betas8_base1}, \ref{fig:slices_betas4_base1} and \ref{fig:slices_betas8_base1}, have a topology similar to the one seen in the baseline, Figure \ref{fig:3Dbase1}. The coolant channels are, for the most part, attached to the symmetry plane. The upper channel of the design optimized with $\bini=4$ is detached from the symmetry plane for a little distance. This increases the contact surface between the fluids and the solid, but probably also has an influence on the heat transfer coefficient. 

When using the baseline design with four channels, Figure \ref{fig:3Dbase2}, and $\bini>1$, the optimized designs seen in Figures  \ref{fig:structure_betas4_base2}, \ref{fig:structure_betas8_base2}, \ref{fig:slices_betas4_base2} and \ref{fig:slices_betas8_base2}, have coolant fluid domain topologies, that also are preserved to a certain extent. The coolant fluid channel paths are, however, modified significantly in comparison to the baseline design. This also affects the flow of the cooled fluid, which the optimized structure disturbs more than what is observed in the baseline design. This higher disturbance of the cooled flow probably also is of benefit to the heat transfer.

Table \ref{tab:3d_obj_compare} shows the obtained heat transfer in the different optimized designs. All optimized designs perform better, than both baselines, whose heat transfers also are seen in Table \ref{tab:3d_obj_compare}. The relative improvements, compared to the better 4 channel baseline, range from $74.8\%$ (starting from $\bini=1$, $\xi=0.5$ and using one channel design pressure drops) to $113\%$ (starting from $\bini=4$ with the four channel baseline). Furthermore, with both baseline designs, it is seen that the designs optimized with the initial projection sharpness $\bini=4$ are the ones performing best. 
\begin{table}[htbp]
    \centering
    \caption{Comparison of the thermal energy flux out of the cold fluid. It is seen that there, in all cases, is a considerable improvement of the baseline design. The designs optimized with the 4 channels baseline design always perform better. Furthermore, with both baseline designs, starting with $\beta=4$ yields the best design.}
    \label{tab:3d_obj_compare}
    \resizebox{\linewidth}{!}{\begin{tabular}{llrr}
        & & 1 channel & 4 channels \\\hline
        \multicolumn{2}{l}{Baseline design} & $4.68$ & $7.27$\\\hline
        \multirow{4}{*}{\rotatebox[origin=c]{90}{Optimized}}&$\xi_\mathit{initial}=0.5$, $\bini=1$ & $12.71$ & $13.69$\\
       &$\bini=1$ & $13.31$ & $13.80$ \\
       &$\bini=4$ & $13.44$ & $15.49$\\
       &$\bini=8$ & $13.34$ & $15.16$
    \end{tabular}}
\end{table}
Redimensionalizing the best optimized design, the transferred thermal power is obtained, to be $P = 89.9\;\mathrm{W}$. Likewise, the best baseline (4 channels), has a transferred thermal power of $P = 42.2\;\mathrm{W}$.

\section{Discussion and conclusion}
\label{sec:conclusion}
In this paper, a new approach for topology optimization of two fluid heat exchangers is presented. The formulation, based on a single design variable field, guarantees the presence of a solid interface between the two fluid phases based on an erosion-dilation identification method. The two mass transfer and one heat transfer problems are solved sequentially in the entirety of the design domain. The optimization maximizes the amount of transferred heat with maximum pressure drop constraints on both fluids. The methodology is applied to two numerical examples.

In the two-dimensional counter-flow example where the design freedom is limited, it is seen that the channels are made narrower, as the allowable pressure drop is increased. This provides an excellent agreement between the amount of transferred heat and the maximum allowable pressure drop obtained using Poiseuille flow and the $\epsilon-\mathit{NTU}$ method. This indicates that the improvement in heat transfer, obtained with the topology optimization process introduced here, can be related to a theoretically derived optimum (for a "one-dimensional" optimization) - at least for this simple case.

In the three-dimensional example, simple shell-and-tube-like baseline designs were significantly enhanced, with up to a $113\%$ improvement compared to the best-performing baseline. In the obtained results, the importance of the flow field in the initial design was highlighted. This relates to the computation of the sensitivities in the early design iterations, which are crucial to the later optimization process. In the optimized designs, microvilli-like features are observed, which enhance both the surface area, but also perturb the flow near them.

The novel and intricate designs appearing when utilizing the presented methodology have been challenging to export to a CAD-based engineering analysis environment. More research into the transformation from density-based representations to CAD/spline basis are needed. In order to evaluate the performance of the obtained designs, a post-evaluation in a commercial software using a segregated approach would indeed be relevant. However, we have so far not been successful in developing a methodology able to transfer the complex optimized designs for further processing. Thus, the obtained designs may at present remain inspirational for future heat exchanger designs.

In this first work on systematic heat-exchanger design with well-defined wall-thickness using topology optimization, we have focused on method development and providing examples with ultimate design freedom, which amongst others resulted in the interesting appearance of microvillies. Such intricate geometrical details will probably not be relevant for practical systems, but give valuable insight. As discussed above, such details may also be difficult to handle geometrically - also manufacturing-wise, even with advance additive manufacturing techniques. To hinder such complexity and ensure manufacturability, we may in future work include more restrictive geometry constraints like larger length-scale, overhang or extrusion constraints, keeping in mind that every constraint imposed will decrease the achievable heat exchanger efficiencies.

\section*{Acknowledgements}
The authors gratefully acknowledge the financial support of the Villum Foundation through the Villum Investigator project InnoTop.


\FloatBarrier
\appendix
\section{Weak formulations and stabilization parameters}
\label{app:matrix_stab}
\subsection{Mass transfer}
\label{app:NSassembly}
The weak form of the Navier-Stokes equations (\ref{eq:NS},\ref{eq:masscons}) is to be found in the follwing. For this purpose, the finite dimensional trial- and test function spaces for the velocity, $\mathcal{S}_u$ and $\mathcal{V}_u$, and for the pressure, $\mathcal{S}_p$ and $\mathcal{V}_p$, respectively, are introduced. The trial functions $u_i\in\mathcal{S}_u$ and $p\in\mathcal{S}_p$ shall be found such that the test functions $\forall w_i\in\mathcal{V}_u$ and $\forall q\in\mathcal{V}_p$. After multiplication by the test functions, the expressions are integrated over the volume $\Omega$.

After integration by parts with the assumption of no outward surface traction, the weak form yields:
 
\begin{equation}
\label{eq:theory_NSweak}
\begin{split}
    \mathscr{R}^u &= \int_\Omega w_iu_i\diff{u_j}{x_i}dV + \int_\Omega \frac{1}{Re}\diff{w_i}{x_j}\left(\diff{u_i}{x_j}+\diff{u_j}{x_i}\right)dV\\
    &\qquad-\int_\Omega\diff{w_i}{x_i}PdV+\int_\Omega w_i\alpha u_idV=0\\
    \mathscr{R}^p&=\int_\Omega q\diff{u_i}{x_i}dV=0
\end{split}
\end{equation}

As discussed in Section \ref{sec:FEM}, the residual functionals are stabilised, by adding both SUPG and PSPG stabilisations, $\mathscr{F}^\delta$ and $\mathscr{F}^\epsilon$, respectively:

\begin{equation}
\label{eq:theory_stab_func}
\begin{split}
    \tilde{\mathscr{R}^u}&=\mathscr{R}^u+\mathscr{F}^\delta=0\\
    \tilde{\mathscr{R}^p}&=\mathscr{R}^p+\mathscr{F}^\epsilon=0
\end{split}
\end{equation}

The weak SUPG- and PSPG stabilisation terms are given as a function of $R^u_i$, the residual from the strong form Navier-Stokes formulation \cite{Alexandersen2014}:

\begin{align}
    \mathscr{F}^\delta&=\sum_{e=1}^{N_e}\int_{\Omega_e}\tau_{SU}u_i\diff{w_j}{x_i}R_j^udV\label{eq:theory_SUPG}\\
    \mathscr{F}^\epsilon&=\sum_{e=1}^{N_e}\int_{\Omega_e}\tau_{PS}\diff{q}{x_j}R_j^udV\label{eq:theory_PSPG}
\end{align}

The stabilization factors, $\TSU$ and $\TPS$, used for the mass transfer problems SUPG and PSPG stabilization, respectively, are the same. It is almost identical to the one used by \cite{Alexandersen2016}:
 \begin{equation}
    \TSU=\TPS= \tau = \left(\frac{1}{\tau_1^r}+\frac{1}{\tau_3^r}+\alpha^r\right)^{-\frac{1}{r}}
\end{equation}
where the power in the minimum function is set to $r=2$. The factors are given as:
\begin{align}
    \tau_1&=\frac{4h_e}{\left|\left|\ue\right|\right|_2}\\
    \tau_3&=\frac{h_e^2\mathit{Re}}{4}
\end{align}

The derivatives of the stabilization parameter with respect the velocity state and to the design variable, used for the computation of the Jacobian matrix and of the sensitivities, are given as:
\begin{align}
\label{app:eq:taudu}
    \diff{\tau}{\ue}&=\frac{1}{\tau_1^{3}}\left(\frac{1}{\tau_1^2}+\frac{1}{\tau_3^2}+\alpha^2\right)^{-\frac{3}{2}}\left(-\tau_1\left(\ue^\intercal\ue\right)^{-1}\ue^\intercal\right)\\
    \diff{\tau}{\xi}&=-\alpha\left(\frac{1}{\tau_1^2}+\frac{1}{\tau_3^2}+\alpha^2\right)^{-\frac{3}{2}}\diff{\alpha}{\xi}
\end{align}

\subsection{Heat transfer}
\label{app:CDassembly}
The weak form of the heat transfer problem from \eqref{eq:CD_ND} is obtained by introducing the trial- and test function spaces, $\mathcal{S}_T$ and $\mathcal{V}_T$, respectively. The finite element problem hence translates to finding a temperature field (trial function) $T\in\mathcal{S}_T$, such that the test function $\forall v\in\mathcal{V}_T$. After integration by part, and assuming no outward surface heat flux (i.e. the domain is insulated), the variational formulation yields.

\begin{equation}
    \label{eq:theory_weakCD}
    \mathscr{R}^T=\int_\Omega v\sum_{\gamma=1}^\mathit{NF} \left(\mathit{Pe}_s^\gamma u_i^{\gamma}\right)\diff{T}{x_i}dV+\int_\Omega C_k\diff{v}{x_i}\diff{T}{x_i}dV=0
\end{equation}

The residual is stabilised with a SUPG stabilisation scheme, $\mathscr{F}^\zeta$:

\begin{equation}
\label{eq:theory_CDres}
    \tilde{\mathscr{R}^T}=\mathscr{R}^T+\mathscr{F}^\zeta=0
\end{equation}

As in the mass transfer, the weak stabilization term is given as a function of the strong for residual  $R_i^T$, obtained from \eqref{eq:CD_ND}, as well as the convection term, $\sum_{\gamma=1}^\mathit{NF} \left(\mathit{Pe}_s^\gamma u_i^{\gamma}\right)$:

\begin{equation}
\label{eq:theory_CDSUPG}
    \mathscr{F}^\zeta=\sum_e^{N_e}\int_{\Omega_e}\tau_{SU_T}\sum_{\gamma=1}^\mathit{NF} \left(\mathit{Pe}_s^\gamma u_i^{\gamma}\right)\diff{v}{x_i}R^T_idV
\end{equation}

The factor for the SUPG stabilization of the heat transfer, $\TSUT$, is given as used by \cite{Alexandersen2016}: 
\begin{equation}	
    \TSUT = \left(\frac{1}{\tau_\mathit{1,T}^r}+\frac{1}{\tau_\mathit{3,T}^r}\right)^{-\frac{1}{r}}
\end{equation}
where the power in the min function is set to $r=2$. The factors are given as:
\begin{align}
    \tau_\mathit{1,T}&=\frac{4h_e}{\left|\left|\ueT\right|\right|_2}\\
    \tau_\mathit{3,T}&=\frac{h_e^2}{4C_k}
\end{align}

The velocity and design variable derivatives of the stabilization parameters are:
\begin{align}
    \diff{\TSUT}{\ueb}&=-\TSUT^3\frac{\Peb}{16h_e^2}\ueT\\
    \diff{\TSUT}{\xi}&=\frac{1}{\tau_\mathit{3,T}^{3}}\left(\frac{1}{\tau_\mathit{1,T}^2}+\frac{1}{\tau_\mathit{3,T}^2}\right)^{-\frac{3}{2}}\left(-\frac{h_e^2}{4C_k^2}\right)\diff{C_k}{\xi}
\end{align}

\section{Notes on sensitivity analysis}
\label{app:sens}
\subsection{Adjoint sensitivity analysis}
\label{app:adjoint}
In order to find the sensitivities of a function, the Lagrangian function, with the arbitrary Lagrangian multipliers, as seen in Equation \eqref{eq:lagrangian}, is set up. Differentiating this function with respect to the design variable, and invoking the chain rule on all terms gives:
\begin{equation}
\label{app:eq:lag_chain}
\begin{split}
    \ddiff{\mathcal{L}}{\xi}=\diff{\Phi}{\xi}+&\sum_{s=\{\mathit{F1},\;\mathit{F2},\;\mathit{T}\}}\left(\diff{\Phi}{\mathbf{s}}\ddiff{\mathbf{s}}{\xi}+\right.\\
    &\bm{\lambda}_{s}^\intercal\diff{\mathbf{R}_{s}}{\xi}+\bm{\lambda}_{s}^\intercal\diff{\mathbf{R}_{s}}{\mathbf{u_1}}\ddiff{\mathbf{u_1}}{\xi}+\\
    &\left.\bm{\lambda}_{s}^\intercal\diff{\mathbf{R}_{s}}{\mathbf{u_2}}\ddiff{\mathbf{u_2}}{\xi}+\bm{\lambda}_{s}^\intercal\diff{\mathbf{R}_{s}}{T}\ddiff{T}{\xi}\right)
\end{split}
\end{equation}
where the summation over $s$ signifies, that the operation is repeated for both mass transfer- and the heat transfer state. As the two mass transfer states are mutually independent, and the coupling between the two mass transfers is weak, the corresponding terms in \eqref{app:eq:lag_chain} can be left out. The Lagrangian multipliers, being arbitrary allows to set  certain terms to zero:
\begin{equation}
\begin{split}
\ddiff{\mathcal{L}}{\xi}=\diff{\Phi}{\xi}+&\bm{\lambda}_{F1}^\intercal\diff{\mathbf{R}_{F1}}{\xi}+\bm{\lambda}_{F2}^\intercal\diff{\mathbf{R}_{F2}}{\xi}+\bm{\lambda}_{T}^\intercal\diff{\mathbf{R}_{T}}{\xi}\\
+&\underbrace{\left(\diff{\Phi}{\mathbf{u_1}}+\bm{\lambda}_{F1}^\intercal\diff{\mathbf{R}_{F1}}{\mathbf{u_1}}+\bm{\lambda}_{T}^\intercal\diff{\mathbf{R}_{T}}{\mathbf{u_1}}\right)}_{=0}\ddiff{\mathbf{u_1}}{\xi}\\
+&\underbrace{\left(\diff{\Phi}{\mathbf{u_2}}+\bm{\lambda}_{F2}^\intercal\diff{\mathbf{R}_{F2}}{\mathbf{u_2}}+\bm{\lambda}_{T}^\intercal\diff{\mathbf{R}_{T}}{\mathbf{u_2}}\right)}_{=0}\ddiff{\mathbf{u_2}}{\xi}\\
+&\underbrace{\left(\diff{\Phi}{T}+\bm{\lambda}_T^\intercal\diff{\mathbf{R}_T}{T}\right)}_{=0}\ddiff{T}{\xi}
\end{split}
\end{equation}
which hence allows for the elimination of the difficult terms $\ddiff{\mathbf{u_1}}{\xi}$, $\ddiff{\mathbf{u_2}}{\xi}$ and $\ddiff{T}{\xi}$, by solving the adjoint problems, outlined in Equations (\ref{eq:sens_lT}-\ref{eq:sens_lF2}). The sensitivities are then found using the Lagrangian multipliers computed by solving the adjoint problems, as seen in Equation \eqref{eq:sens_2f}.

\subsection{Chain rule projection of sensitivities}
\label{app:chainrule}
The sensitivities of the objective function, and of the constraints, are found with respect to $\xiI$ and $\xiII$ , as described in Section \ref{sec:optsens} and \ref{app:adjoint}. These sensitivities $\diff{\Phi}{\xiI}$ and $\diff{\Phi}{\xiII}$, are projected back to $\diff{\Phi}{\xi}$ using the chain rule:
\begin{equation}
\label{app:eq:chainrule_def}
\diff{\Phi}{\xi}=\diff{\Phi}{\xi}+\ddiff{\Phi}{\xiI}\diff{\xiI}{\xi}+\ddiff{\Phi}{\xiII}\diff{\xiII}{\xi}
\end{equation}
The sensitivities with respect to the eroded and dilated variables are projected back to the design variable $\xi$ with steps corresponding to the one of the erosion dilation process from Section \ref{sec:erodila}. Keeping in mind that $\diff{\xiI}{\xihtI}=-1$, the chain-rule terms are given as:
\begin{align}
    \diff{\xiI}{\xi}&=-\diff{\xihtI}{\xip}\diff{\xip}{\xih}\diff{\xih}{\xit}\diff{\xit}{\xi}\\
    \diff{\xiII}{\xi}&=\diff{\xihtII}{\xip}\diff{\xip}{\xih}\diff{\xih}{\xit}\diff{\xit}{\xi}
\end{align}

The chain rule terms $\diff{\xiht_\gamma}{\xi}$ and $\diff{\xih}{\xi}$ are found by differentiation of the smooth Heaviside projection from Equation \eqref{eq:topopt_heavi}:
\begin{equation}
    \label{eq:topopt_heavisens}
    \diff{\hat{\xi}}{\tilde{\xi}}=\beta\frac{1-\tanh^2(\beta(\tilde{\xi}-\eta))}{\tanh(\beta\eta)+\tanh(\beta(1-\eta))}
\end{equation}
The filtering operation terms from the chain rule, $\diff{\xiht}{\xih}$ and $\diff{\xit}{\xi}$, are equivalent to filtering the sensitivities through the PDE filter, as was done with $\xi$ to obtain $\xit$ and with $\xih$ to obtain $\xiht$ \cite{Lazarov2011}.


\bibliographystyle{elsarticle-num-names}
\bibliography{bib.bib,bibjoe.bib}

\end{document}